\documentclass[epj,numbook]{SVJour}
\usepackage{amssymb,epsf}

\newcommand{\erf}{\mbox{erf}}
\newcommand{\cP}{{\cal P}}
\newcommand{\D}{\displaystyle}
\newcommand{\xis}{\xi_{\perp}}
\newcommand{\xip}{\xi_{\parallel}}
\newcommand{\eff}{{\rm eff}}
\newcommand{\rb}[1]{\raisebox{1.5ex}[-1.5ex]{#1}}

\newcommand{\pic}[3]
{
 \begin{figure}[bt]
   \begin{center}
     \leavevmode
     \epsfxsize=#2
     \epsfbox{#1.ps}
   \end{center}
   \caption{\label{#1} #3}
 \end{figure}
}

\begin{document}

\title{Critical behavior of interacting surfaces with tension}

\dedication{Dedicated to Johannes Zittartz on the occasion of
  his 60th birthday.}

\author{
  Andreas Volmer \inst{1} \and 
  Udo Seifert \inst{2} \and 
  Reinhard Lipowsky \inst{2}
}

\institute{
  Institut f\"ur theoretische Physik, Universit\"at K\"oln, 
  Z\"ulpicher Str. 77, D-50937 K\"oln, Germany 
\and
  Max-Planck-Institut f\"ur Kolloid- und Grenzfl\"achenforschung, 
  Kantstra{\ss}e 55, D-14513 Teltow-Seehof, Germany
}

\date{\today}

\abstract{
  Wetting phenomena, molecular protrusions of lipid bilayers and
  membrane stacks under lateral tension provide physical examples for
  interacting surfaces with tension. Such surfaces are studied
  theoretically using functional renormalization and Monte Carlo
  simulations.  The critical behavior arising from thermally--excited
  shape fluctuations is determined both for global quantities such as
  the mean separation of these surfaces and for local quantities such
  as the probabilities for local contacts.
}

\maketitle

\section{Introduction}

\label{Sec:1} 

The surfaces considered here are  interfaces and membranes
which are governed by tension. An interface or domain wall
which represents the stable contact region between two bulk phases of
matter is always characterized by  a finite interfacial tension, see,
e.g., \cite{rowl82}.
Flexible membranes, on the other hand, are sheets of amphiphilic
molecules such as lipid bilayers and are usually controlled by
curvature and bending rigidity, see, e.g., \cite{liposack}. However, these
membranes may be subject to an applied lateral tension arising from external
forces or constraints. In addition, lipid bilayers exhibit molecular
protrusions on small scales which are also governed by an effective
tension.

In this paper, we study the interactions of two or several such
surfaces and focus on the renormalization of these interactions
arising from thermally--excited shape fluctuations.  Physical examples
are provided by (i) Wetting layers which are bounded by two
interfaces; for reviews, see \cite{lipo78,diet88,fish89a,schi90}. In
this case, the interactions between these two interfaces are
renormalized by capillary waves \cite{lipo39}; (ii) Lipid bilayers at
separations which are small compared to the bilayer thickness. Such a
situation arises for stacks of bilayers in the presence of a
relatively large external pressure acting on these stacks. In this
case, the interactions between adjacent bilayers are renormalized by
molecular protrusions  -- thermally excited displacements
of neighbouring lipid molecules which change the
area of the lipid-water interface \cite{lipo105+lipo108}; and (iii)
Swollen 
bunches or stacks of membranes at separations which are large compared
to the thickness of these membranes.  This corresponds to the limit of
relatively small external pressures acting on the membrane bunch or
stack. Previous theoretical work on such bunches has focused on
tensionless membranes as reviewed in Ref. \cite{lipo120}. In contrast,
we will be concerned here with membranes under lateral tension.

The paper is organized as follows.  In Section \ref{Sec:2}, we will
define the theoretical models appropriate for (i) complete wetting,
(ii) protrusion forces between bilayers, and (iii) membranes under
lateral tension and explain that these models are equivalent to each
other if the various parameters are identified in an appropriate
way. In Section \ref{Sec:3}, we define the global and local critical
effects which we will study in the remainder of the paper. One
critical effect which has not been studied previously in a systematic
way is the behavior of the probability for local contacts between the
surfaces. Our results for two surfaces with hard wall and soft wall
interactions are described in Sections \ref{Sec:4} and \ref{Sec:5},
respectively.

In order to simplify the presentation, we will explicitly discuss our results
in Sections \ref{Sec:4} and \ref{Sec:5} using the terminology of
molecular protrusions for lipid 
bilayers. However, if one makes the appropriate identification of parameters as
explained in Section \ref{Sec:2},  these results can be directly
applied both to  complete wetting and to membranes under lateral tension.


\section{Theoretical framework}

\label{Sec:2} 

\subsection{Wetting phenomena}

Consider an interface between two bulk phases, say $\alpha$ and
$\gamma$, and let us change a thermodynamic field, such as
e.g. temperature or pressure, in order to move the system towards a
triple point where a third phase, $\beta$, can coexist with the two
phases $\alpha$ and $\gamma$. Then, in thermal equilibrium, a layer of
the $\beta$ phase may appear in the $\alpha \beta$ interface. As one
comes closer and closer to the triple point, the thickness of this
intermediate layer may continuously grow: this is the case of {\it
complete wetting}. On the other hand, no layer may appear in the
$\alpha \beta$ interface or the thickness of this layer may saturate
and remain finite as the triple point is attained: this is the case of
{\it incomplete wetting}.

When the $(\alpha \gamma)$ interface contains a wetting layer of
$\beta$ phase, it splits up into an $(\alpha \beta)$ and a $(\beta
\gamma)$ interface. The mean separation, $\ell \ge 0$, of these two
interfaces is equal to the thickness of the wetting layer. Likewise,
the excess free energy of the layer can be regarded as an effective
interaction between these interfaces. Here, we will focus on the
situation where the interaction potential between these two interfaces
is purely repulsive and short--ranged. An example is provided by
complete wetting in 3--dimensional lattice gas or Ising models.

\subsubsection{Solid--on--solid models for  wetting}

Lattice gas or Ising models on a cubic lattice with a planar surface
presumably provide the simplest models for wetting in three dimensions
even though the corresponding phase diagram is already quite complex,
see, e.g., \cite{bind88}.  Wetting occurs at or close to two--phase
coexistence provided the bulk of the system is in one phase, say in
the spin--up phase (corresponding to the dilute vapor phase in the
lattice gas) whereas the surface of the semi--infinite system favors
the other phase (corresponding to the dense liquid phase in the
lattice gas). In such a situation, one has a thin wetting layer of
spin--down phase which is bounded (i) by the interface between this
layer and the bulk phase and (ii) by the planar surface of the
semi--infinite system.

At low temperatures, the interface is essentially flat. As one
increases the temperature, typical excitations consist of islands on
this flat interface.  These excitations can be studied in the
framework of so--called solid--on--solid models in which the planar
surface of the semi--infinite system is described by a square lattice
with lattice sites $x_i$ and lattice constant $a$, and the local
separation of the interface from this surface is described by the
separation field $l_i \equiv l(x_i)$.  The configurational energy or
effective Hamiltonian of this separation field is given by
\begin{equation}
  \label{SOSham}
  {\cal H} \{l\}  = \sum_{\langle ij \rangle} (2 J /a)
  |l_i-l_j| + \sum_i a^2 V(l_i)
\end{equation}
where ${\langle ij \rangle}$ indicates a summation over all nearest
neighbors, $J$ is the nearest neighbor coupling constant in the Ising
model, and $V(l)$ denotes the effective potential acting on the
interface. \cite{footnote1} For complete wetting, this potential has
the simple form
\begin{equation}
  \label{Vhw+Hl}
  V(l) = V_{hw}(l) + H l
\end{equation}
where the hard--wall potential $V_{hw}(l)$ is defined by $V_{hw}(l)=
0$ for $l>0$ and $V_{hw}(l) = \infty$ for $l<0$ and $H$ is the bulk
magnetic field in the original Ising model.  At temperature $T$
(measured in energy units), the statistical weight for the
configuration $l_i = l(x_i)$ is given by the Boltzmann factor $\sim
{\rm exp} [- {\cal H} \{ l \} /T]$.

\subsubsection{Gau{\ss}ian interface models for  wetting}

The model as given by (\ref{SOSham}) and (\ref{Vhw+Hl}) has been
studied by Monte Carlo simulations. \cite{lipo63,lipo87} In these
simulations, the mean separation of the two surfaces was determined in
the limit of small magnetic field $H$.  It was found that this
quantity exhibits the same critical behavior in the discrete
solid--on--solid model as in the corresponding Gau{\ss}ian interface
model.  In its discrete version, the latter model is defined by the
effective Hamiltonian
\begin{equation}
  {\cal H} \{l\}  = \sum_{\langle ij \rangle} \frac{1}{2} \Sigma
  (l_i-l_j)^2 + \sum_i a^2 V(l_i)
\end{equation}
where $\Sigma$ represents the interfacial tension (or stiffness) of
the fluctuating interface on large scales. In the continuum limit,
this leads to 
\begin{equation}
\label{GaussContHam}
{\cal H} \{  l\} = \int d^2 x \left\{ \frac{1}{2} \Sigma
(\nabla l)^2 +  V(l) \right\}
\end{equation}
where both $x$ and $l$ are   continuously varying coordinates. The
latter model has been studied by functional renormalization as
discussed in Ref. \cite{lipo105+lipo108}.

The critical behavior of the mean separation $\langle l \rangle$ is
identical in the solid--on--solid and in the Gau{\ss}ian model
provided one identifies the interfacial tension with
\begin{equation}
\Sigma = c_\Sigma (2J/a)^2 /T  \qquad {\rm with} \quad c_\Sigma \simeq 2.6
\end{equation}
for the square lattice.

In the present work, we will address the question to what extent this
correspondence between the two models is also valid for other
quantities such as, e.g., the probability for local contacts. In fact,
we will see that the latter quantity has the same behavior for hard
wall interactions but exhibits different behavior for soft
wall interactions, see Sect.~5 below.

\subsection{Molecular protrusions of bilayers}

Now, let us consider a rather different system consisting of several
lipid bilayers. It has been well--established by many experiments that
such bilayers in aqueous solution experience strong repulsive forces
at small separations of the order of 1 nm. \cite{rand89} It was
originally thought that this short--ranged repulsion represents a
hydration effect and reflects the perturbed water structure in front
of the polar head group \cite{marc76}. More recently, it has been
shown, however, that a similar repulsive force can also arise from the
molecular roughness of the lipid--water interfaces.

Lipid bilayers consist of rod--like molecules with a hydrophilic head
group and usually two hydrophobic tails. These rods are packed in such
a way that the hydrophilic head groups shield the tails from the
surrounding water. The corresponding lipid/water interface is
roughened by thermally--excited fluctuations as has been observed in
computer simulations \cite{egbe88,past94} and has been deduced from
scattering experiments \cite{koen92,wien92,mcin93}. These thermal
fluctuations correspond to protrusions, i.e, to relative displacements
or deformations of the lipid head groups. Since these protrusions
will, in general, change the surface area of the lipid--water
interface, they are governed by an effective tension.

The repulsive force arising from the protrusions of single molecules
was calculated in Ref. \cite{isra90}. The effect of collective
protrusions which involve whole groups of lipid molecules was first
studied in Ref. \cite{lipo105+lipo108} within the framework of
discrete and continuum interface models.

\subsubsection{Models for collective protrusions}

Consider one such lipid/water interface formed by the head groups of
the lipid molecules which is located in front of a planar surface or
wall. The head group at position $x_i$ has the local separation $l_i $
from the wall. If the lipid/water interface is flat, its separation
from the wall is constant and all $l_i $ are equal. Thermally--excited
protrusions roughen this flat interface and lead to
position--dependent configurations $l_i = l(x_i)$.

The configurational energy of these protrusions consists of two
contributions.  The first contribution is given by the excess free
energy arising from the relative displacements of the molecules which
is governed by the interfacial tension $\Sigma_o$. The latter quantity
represents the free energy per unit area for the interface between the
nonpolar part of the molecule and the water. In the discrete model to
be defined, each molecule is supposed to have $n_{nn}$ nearest
neighbors. The molecular shape is approximated by a rigid column with
circumference $a_o$ and cross-sectional area $A_o$. For $n_{nn} = 6$,
one has hexagonal columns on a triangular lattice with lattice
constant $a= a_o/ 2 \sqrt{3}$ and cross--sectional area $A_o =
\sqrt{3} a^2 / 2$.

The second contribution to the configurational energy of the protrusions
represents the excess free energy arising from the interaction of the
lipid/water interface with the planar surface or wall. The corresponding
interaction potential  $V(l)$ has the generic form
\begin{equation}
\label{Vexplicit}
V(l) = V_{hw} (l) + V_{hy} \exp[- l / l_{hy}] + Pl
\qquad .
\end{equation}
The first term of $V(l)$ is again the hard wall repulsion which
ensures that the two surfaces cannot penetrate each other.  The second
term in (\ref{Vexplicit}) represents the hydration interaction arising
from the perturbed water structure in front of the lipid head groups;
it is parameterized by a certain decay length $l_{hy}$ and a certain
amplitude $V_{hy}$. The last term $Pl$ is the excess free energy
arising from the external pressure $P$.  For finite $P$, the
lipid/water interface will have a finite mean distance from the wall
and will undergo an unbinding transition in the limit of small $P$.

If one combines both contributions, one obtains
the configurational energy or effective Hamiltonian
\cite{lipo105+lipo108}
\begin{equation}
\label{ProtrusionHam}
{\cal H} \{ l \}= \sum_{<ij>} (a_o \Sigma_o/n_{nn} ) \\\
| l_i - l_j | + \sum_i \; A_o V(l_i)
\end{equation}
for the collective protrusions described by the configuration $l_i = l(x_i)$.
Comparison with the solid--on--solid model  (\ref{SOSham}) shows that both
models are completely equivalent.

The model as given by (\ref{ProtrusionHam}) was studied by Monte Carlo
simulations for a 
triangular lattice with $n_{nn} = 6$; in addition, the corresponding
Gau{\ss}ian model was studied by functional renormalization.
\cite{lipo105+lipo108} It was again found that the critical behavior
of the mean separation is the same in both models provided one chooses
the effective interfacial tension 
\begin{equation}
\label{EffectiveTension}
  \Sigma \equiv \Sigma_{pr} = c_\Sigma (a_o \Sigma_o/n_{nn})^2  /T .
\end{equation}
For the triangular lattice with $n_{nn} = 6$ nearest neighbors,
the Monte Carlo simulations in Refs. \cite{lipo105+lipo108} lead to the
estimate $c_\Sigma \simeq 2.4$ which is different from but close to the value
$c_\Sigma \simeq 2.6$ as obtained for the square lattice.

\subsection{Membranes under lateral tension}

On sufficiently large scales, flexible membranes which do not
experience any external force or constraint are governed by their bending
rigidities and curvature energies, see the various reviews in Ref.
\cite{liposack}. On the other hand, the presence of such forces or constraints
tends to induce lateral tensions which reduces the bending undulations of the
membranes
\cite{broc75a,helf84}. One example is provided by a closed vesicle which is
inflated by an osmotic pressure difference
$\Delta P = P_{in} - P_{ex}$. The surface of this vesicle will attain an
essentially spherical shape with radius
$R_{ve}$ and the corresponding tension $\Sigma$ satisfies the Laplace
equation $\Sigma =  R_{ve} \Delta P/2$.

If the vesicle is unilamellar, i.e., if it is bounded by a single membrane,
this membrane is subject to the tension $\Sigma$. If the vesicle is
multi--lamellar, i.e., if the vesicle surface consists of $N$ closely
packed membranes, each membrane should experience a lateral tension of the
order of $\Sigma/N$.

It is also possible to apply a controlled suction pressure $\Delta P$ to the
vesicle by micropipet aspiration \cite{evan87,evan90}. In this way, one can
directly control the lateral tension $\Sigma$. For a multi--lamellar vesicle,
one then has $N$ interacting membranes under lateral tension. In addition,  the
tense vesicle can also be pushed, via the micropipet, against a planar substrate
or wall. Within the contact region, one then has an oriented stack
interacting with the planar surface.

\subsubsection{Continuum Gau{\ss}ian models}

Now, consider an oriented stack of membranes labeled by $n$ with
 $1 \le n \le N$. Each membrane is characterized by its bending rigidity
$\kappa_n$ and is taken to experience the lateral tension $\Sigma_n$.
For an oriented stack of membranes, the position of membrane $n$ can be
described by the height variables $h_n(x)$  which measures the distance
from a reference plane with coordinate  $x$.
The interaction potential between two neighboring membranes is again
denoted by $V(h_{n+1}-h_n)$. Using this parameterization,
the effective Hamiltonian for the whole stack has the form
\begin{eqnarray}
{\cal H} \{h \}  =
 \int d^{2} {\bf x} 
 &\bigg\{ &
 \sum_{n=1}^{N} \left[
   \frac{1}{2} \Sigma_{n} \left(\nabla h_{n} \right)^{2} +
   \frac{1}{2} \kappa_{n}
   \left(\nabla^{2} h_{n} \right)^{2}  \right]   \nonumber \\
 &&  +\sum_{n=1}^{N-1}  V(h_{n+1} - h_n)  \bigg\}
\end{eqnarray}
For $N=2$, one has a rigidity--dominated regime
for sufficiently small scales and a tension--dominated regime for sufficiently
large scales \cite{liposack,lipo119}. In the latter regime, the bending terms
$\sim \left(\nabla^{2} l_{n} \right)^{2}$ become irrelevant. Such a behavior is
to be expected for general  $N > 2$,
and the critical behavior is then governed by the interplay between the tension
terms and the interaction terms.  In order to study this interplay, it is
convenient to make an orthogonal transformation from  rescaled height variables
$ \sqrt{\Sigma_n} h_n$ to new fields $z_n$ which contain the 'center--of--mass'
coordinate
\begin{equation}
z_N \equiv \sum_{n=1}^{N} \Sigma_n h_n /
\left[\sum_{n=1}^{N} \Sigma_n \right]^{1/2}
\quad .
\end{equation}
The latter coordinate decouples from the other fields $z_n$
with $ n < N$ since it does not enter in the  interaction terms
$\sim V(h_{n+1} - h_n)$.
In this way, one arrives at models for the $N-1$ fields $z_n$
with $n < N$ which are
linear combinations of the $N-1$ separation fields $l_n \equiv h_{n+1} - h_n$.
For $N=2$, one obtains
a model for the single field $z_1 \sim h_2 - h_1$ which is equivalent to the
continuum Gau{\ss}ian model for wetting as given by (\ref{GaussContHam}).

\section{Critical behavior}

\label{Sec:3} 

In the following, we will discuss our results using the terminology which
is appropriate for molecular protrusions of bilayers.
The   potentials $V(l)$
considered here consist of   repulsive hard--wall and soft--wall interactions
balanced by an external pressure term $P l$ as given in (\ref{Vexplicit}).
In the present section, we will define the global and local quantities  which
become critical as the pressure  $P$ goes to zero.

\subsection{Global quantities}

Thus, consider two surfaces governed by tension which are pushed together by
the external pressure $P$. The configuration of these two surfaces is described
by their local separation $l=l(x)$. As
$P$ is decreased to zero, the two surfaces unbind which leads to the divergence
of several length scales. First of all,
the parallel correlation length $\xi_{\|}$ diverges  as
\begin{equation}
\xi_{\|} \sim 1/P^{\nu_{\|}}
\end{equation}
which  defines the critical exponent $\nu_{\|}$.

In addition,  both  the mean
separation $\ell \equiv \langle l \rangle$ of the two surfaces and the roughness
\begin{equation}
\xi_{\bot} \equiv \left\langle (l - \langle l \rangle)^2 \right\rangle^{1/2}
\end{equation}
of the separation field $l$ are singular for small $P$. For protrusions, which
are governed by  the effective tension $\Sigma = \Sigma_{pr}$ as in
(\ref{EffectiveTension}), 
the interfacial roughness is related to $\xi_{\|}$ via
\begin{equation}
\label{DefXiBot}
\xi_{\bot} \approx (T / 2 \pi \Sigma)^{1/2} \sqrt{\ln ( \xi_{\|} /
a )}
\end{equation}
where   $a $
denotes the small--scale cutoff as before \cite{lipo39}. This relation
is analogous to the well--known scaling of the roughness of  
a free interface as a function of its linear dimension.
Note that the basic length scale
which sets the size of the interfacial roughness
is given by the protrusion length
\begin{equation}
\label{DefLpr}
l_{pr} \equiv (T/2\pi \Sigma )^{1/2} \quad .
\end{equation}

For three surfaces, one has two separation fields $l_1 = h_2 - h_1$ and
$l_2 = h_3 - h_2$. In this case, one convenient choice is to use the
two fields $d_1 \sim l_1 + l_2$, which is   the thickness of the whole
bunch, and $d_2 \sim l_1 - l_2$.

\subsection{Local contacts}

As the surfaces unbind, the probabilities for local contacts of the
surfaces decay to zero. More precisely, the probability $\cP_{2b}$ for pair
contacts is found to behave as
\begin{equation}
\cP_{2b} \sim   1 / \xi_{\parallel}^{\zeta_2} \sim P^{\nu_2}
\end{equation}
which defines the critical exponents $\nu_2$. This exponent satisfies the
scaling relation
\begin{equation}
\nu_2 \equiv \zeta_2 \nu_{\|}  \qquad .
\end{equation}

For a bunch of three or more interacting surfaces, one may
consider the   probabilities $\cP_{nb}$ for local $n$--contacts.
These quantities have the scaling behavior
\begin{equation}
\cP_{nb} \sim 1 / \xi_{\parallel}^{\zeta_n} \sim P^{\nu_n}
\qquad {\rm with } \quad \nu_n \equiv \zeta_n \nu_{\|} \quad .
\end{equation}

Local contacts have also been  studied   for 1--dimensional lines (or
strings or directed walks) governed by line tension
\cite{lipo102,laes94,lipo117,lipo125} and for tensionless membranes
\cite{lipo125}. In the following sections, we will study these quantities
for two and three interacting surfaces governed by tension.


\section{Two surfaces with hard wall interactions}

\label{Sec:4} 

In this section, we  study the critical behavior of two surfaces
interacting with hard wall interactions $V_{hw}(l)$ with  $V_{hw}(l)= 0$
for $l>0$ and $V_{hw}(l) = \infty$ for $l<0$ as before. Since this
interaction contains no length and no energy scale, it does not
introduce any dimensionfull parameter into the model.

We begin with a brief summary of the results from a functional
renormalization group treatment, which gives predictions for the mean
distance $\ell$ and the roughness $\xi_\bot$ of the
interfaces. Additionally, we consider the pair contact probability
$\cP_{2b}$ and derive predictions for its scaling behavior. Monte
Carlo (MC) simulations are used to test the theoretical results. The
Gau{\ss}ian and the solid--on--solid model are treated
separately.

\subsection{Gau{\ss}ian models}

First, we give a short review of the functional renormalization group
treatment of the Gau{\ss}ian model as defined by (\ref{GaussContHam}),
following \cite{lipo105+lipo108}.
The functional renormalization group is used to integrate out thermal
fluctuations on scales between the microscopic length $a$ and the lateral
correlation length
$\xi_\parallel$, which will be finite for finite values of the
pressure $P$. Technically, this integration is performed in a
one-step nonlinear functional renormalization group with a
rescaling factor $b=\xi_\parallel/a$, using a formalism which is an
extension of Wilson's approximate recursion relation \cite{Wilson71},
and which has been developed in studies of the wetting transition
\cite{lipo39}. This procedure yields an effective interaction
potential on the scale of the lateral and
perpendicular correlation lengths $\xi_\parallel$ and $\xi_\bot$. Now,
on this scale, mean field theory can be used, allowing to set up a self
consistent procedure to determine the mean separation
$\ell\equiv\langle l\rangle$, $\xi_\bot$ and
$\xi_\parallel$ by minimizing the effective potential. This effective
potential is now, on the scale $\xi_\parallel$,
defined as a simple superposition of the effective hard wall potential
and a linear pressure term:
\begin{equation}
  V^{\eff}(l) \equiv V^{\eff}_{hw}(l) + Pl,
\end{equation}
where $V_{hw}^{\eff}(l)=-\tilde v \ln\erf(l/\sqrt2 \xi_\bot)$.
The parameter $\tilde v \simeq T/A_o$ represents the basic energy scale.

Minimization with respect to $l$ leads to
\begin{equation}
  \label{match_1}
  P = \left. -\frac{\partial V_{hw}^{\eff}}{\partial l}
      \right|_{l=\ell} \approx \frac{\sqrt{2}\tilde{v}}{\sqrt{\pi}\xis}
          e^{-\ell^2/2\xis^2}.
\end{equation}
In addition, we have the mean field relation between
$\xi_\parallel$ and the curvature of the effective potential in the
minimum position
\begin{equation}
  \label{match_3}
  \frac{\Sigma}{\xip^2} = \left.
  \frac{\partial^2 V_{hw}^{\eff}}{\partial l^2} \right|_{l=\ell}
\end{equation}
and the relation between the two length scales $\xi_\parallel$ and
$\xi_\perp$ as given in (\ref{DefXiBot}). Solution of this set of
  equations yields a 
relation between the pressure and the mean separation $\ell$ that is given by
\begin{equation}
 \label{P_ell}
 P \approx P_{hw} e^{-\ell/l_{pr}} (l_{pr}/\ell)^{1/4},
\end{equation}
with the protrusion length $l_{pr} = (T/2\pi\Sigma)^{1/2}$ as before.
Likewise, the surface roughness $\xi_{\bot}$ is found to obey the relation
\begin{equation}
\label{P_xis}
  P \approx P_{hw\perp} e^{-2(\xis/l_{pr})^2}.
\end{equation}
The two pressure amplitudes take the form
\begin{equation}
\label{P_Amp}
P_{hw} = \frac{\sqrt{\Sigma \tilde v}}{\pi^{1/4} a} \qquad \mbox{and}
  \qquad P_{hw\perp} = \frac{l_{pr} \Sigma} {2a^2}.
\end{equation}
Inverting the two relations (\ref{P_ell}) and (\ref{P_xis}), one finds
\begin{equation}
\label{ell_P}
\ell \approx l_{pr} \left[\,\ln(P_{hw}/P) - 1/4\;\ln\ln(P_{hw}/P) \,\right]
\end{equation}
and
\begin{equation}
\label{xi_P}
  \xi_{\bot}^2 \approx  l_{pr}^2/2\;\ln(P_{hw\bot}/P)
\end{equation}
in the limit of small $P$.

The mean separation and the correlations lengths represent global
quantities of the interface. In order to introduce a local
quantity, namely the pair contact probability $\cP_{2b}$,
we define the probability distribution
\begin{equation}
\label{PDeltal}
  \cP_\Delta(l) \equiv
  \langle\, \theta(l+\Delta l)-\theta(l) \,\rangle / \Delta l,
\end{equation}
where $\theta(l)$ is the Heaviside step function and $\Delta l$ is a
microscopic length scale. $\cP_\Delta(l)$ gives the probability to
find the surface height $l(x)$, at a given position $x$, in the
interval $l\ldots l+\Delta l$.
Then, the probability $\cP_{2b}$ for pair contacts is given by
\begin{equation}
  \cP_{2b} = \cP_\Delta(l=0).
\end{equation}
In the following, we will omit the subscript $\Delta$, and take $\Delta l$
to be a fixed microscopic length scale $\Delta l$.
If one assumes that $\cP(l)$ is well approximated by a
Gau{\ss}ian distribution which is centered at the mean separation $\ell$ and
has the width $\simeq$  $\xi_{\bot}$, the pair contact
probability is given by
\begin{equation}
\label{P2_ellxi}
  \cP_{2b} \sim \exp[-\ell^2/2 \xi_{\bot}^2] / \xi_\bot \quad .
\end{equation}
Note that in the Gau{\ss}ian case, the full distribution $\cP(l)$ is fully
determined by its first two moments, $\ell$ and $\xi_{\bot}^2$.
Inserting the asymptotic pressure dependence of these two
length scales, as given by (\ref{ell_P})
and (\ref{xi_P}), into (\ref{P2_ellxi}), one immediately obtains
\begin{eqnarray}
  \label{P_2_hw}
  \cP_{2b} &\sim& \exp\left[
    - \ln^2(P_{hw}/P) \;/\; \ln(P_{hw\bot}/P)
  \right]
  \\
  &\sim& P^{\nu_2} \qquad {\rm with} \quad \nu_2 = 1 , \rule{0cm}{2.5ex}
\end{eqnarray}
neglecting confluent logarithmic corrections to the leading scaling
behavior. The contact exponent $\nu_2$, which has been introduced in Sect.
3.2, governs the asymptotic behavior for $P\gg P_{hw}$ and $P\gg P_{hw\bot}$.

The above assumption about the Gau{\ss}ian character of the probability
distribution $\cP(l)$ has been confirmed by
Monte Carlo simulations of the discrete Gau{\ss}ian model. In the
simulations, the discrete form
\begin{equation}
  {\bar{\cal{H}}} \{z\} \equiv 
  \frac{{\cal H}\{z\}}{T} = 
  \sum_{\langle i,j\rangle}
  \frac {(z_i-z_j)^2}{2n_{nn}} 
  + \sum_i (U_{hw}(z_i) + Qz_i)
\end{equation}
of the effective Hamiltonian has been studied,
where we have used the dimensionless quantities
$z\equiv l/l_{sc}$  and $Q\equiv P/P_{sc}$ with
\begin{equation}
  l_{sc}=\sqrt{T/\Sigma n_{nn}}
  \qquad \mbox{and} \qquad
  P_{sc}=\sqrt{T\Sigma n_{nn}}/a^2.
\end{equation}

For all MC simulations, a triangular lattice with 6 nearest neighbours per
site was used, and the simulated area had the form of a
parallelogram. 
Periodic boundary conditions were applied in the
direction parallel to the base line of the parallelogram and in the
direction perpendicular to this line. The linear dimensions in both
directions were always taken to be equal. Typical lattice sizes were
$L_\parallel$=32 and 64, where $L_\parallel$ is the linear
dimension. For optimization of calculation time, the height variables
$z_i$ were discretized; we have checked, however, that the results do
not differ from those obtained when using floating point
variables. Finally, the acceptance rate of the MC step was
continuously controlled by tuning the maximum height variation per
step and site. The length of the MC runs varied between $10^5$ and
$10^6$ MC steps per site, depending on the chosen value of the
external pressure $Q$. 

The statistical errors of the MC results were controlled in the
following way. Starting with a small MC sampling interval $\Delta t$, the
mean values of all observables were measured within $n$ equal
intervals of 
length $\Delta t/n$, giving $n$ values $A_\mu$ with $\mu
=1,\ldots,n$. Usually, we have chosen the value $n=10$.
From these values, the statistical error $\delta A$,
defined as $\delta A^2=\langle\, [\,\frac{1}{n}\sum_{\mu=1}^n (A_\mu -
\langle A\rangle)\,]^2\,\rangle$, was calculated.
Then, $\Delta t$ was multiplied by $n$, and the procedure was
repeated on the larger time scale. In this way, one can plot the
relative statistical error $\delta A/\langle A\rangle$ of the
observable $A$ as a function of $\Delta t$. The results of this
procedure are illustrated in Fig.~\ref{error}. Note that when $\Delta
t$ becomes larger than the relaxation time, the relative error
decreases monotonically. In the MC results presented in this paper, we
have plotted no error bars because the statistical errors were
typically smaller than the data symbols.

\pic{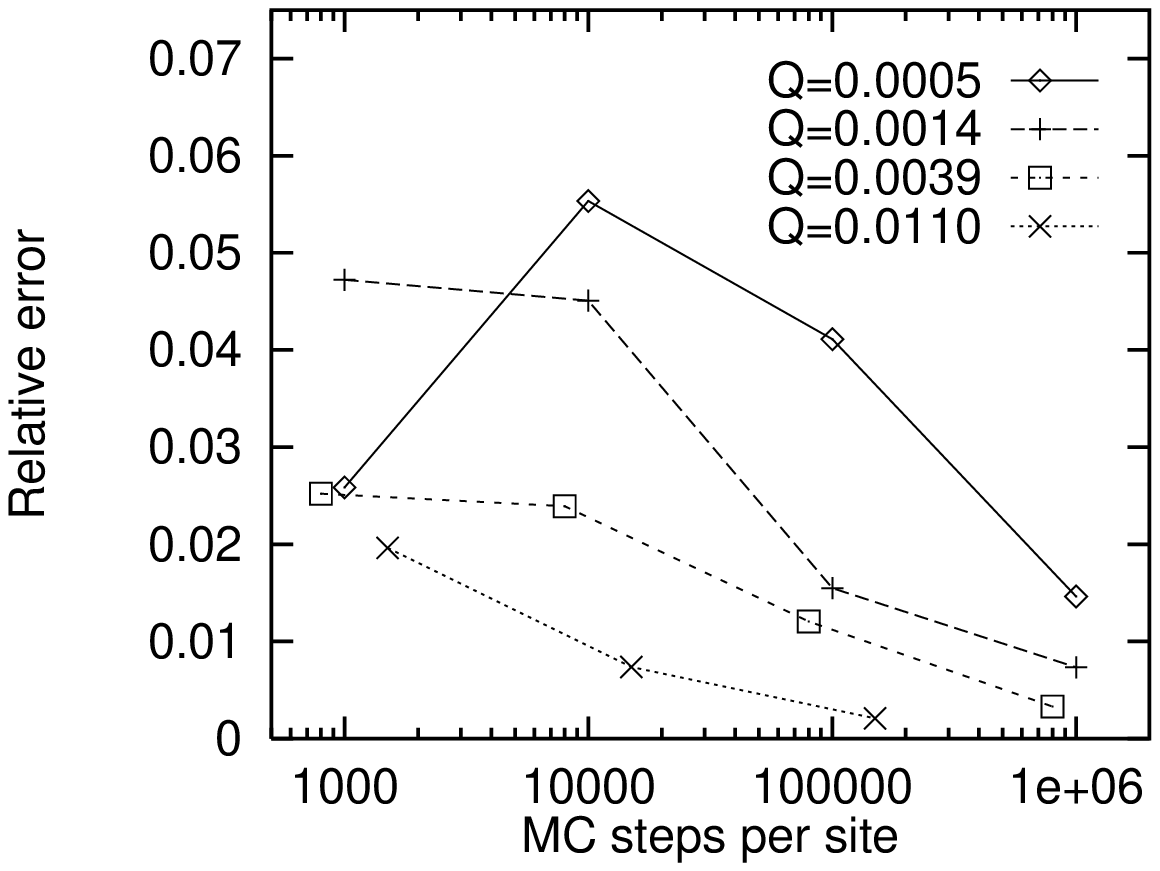}{0.4\textwidth}{Some typical values for the relative error
  of the measured mean distance as a function of the simulation
  time, measured in MC steps per site $\Delta t$. The SOS model with
  one flexible surface fluctuating against a hard wall was
  simulated, with a system of linear dimension $L_{\parallel}=64$.}

MC data of probability distributions $\cP(l)$ for three different values of
the pressure $Q$ are shown in Fig.~\ref{quadww_hist}. These data were
obtained in simulations of one flexible surface pushed against a hard
wall, and are well fitted by Gau{\ss}ian distributions.

\pic{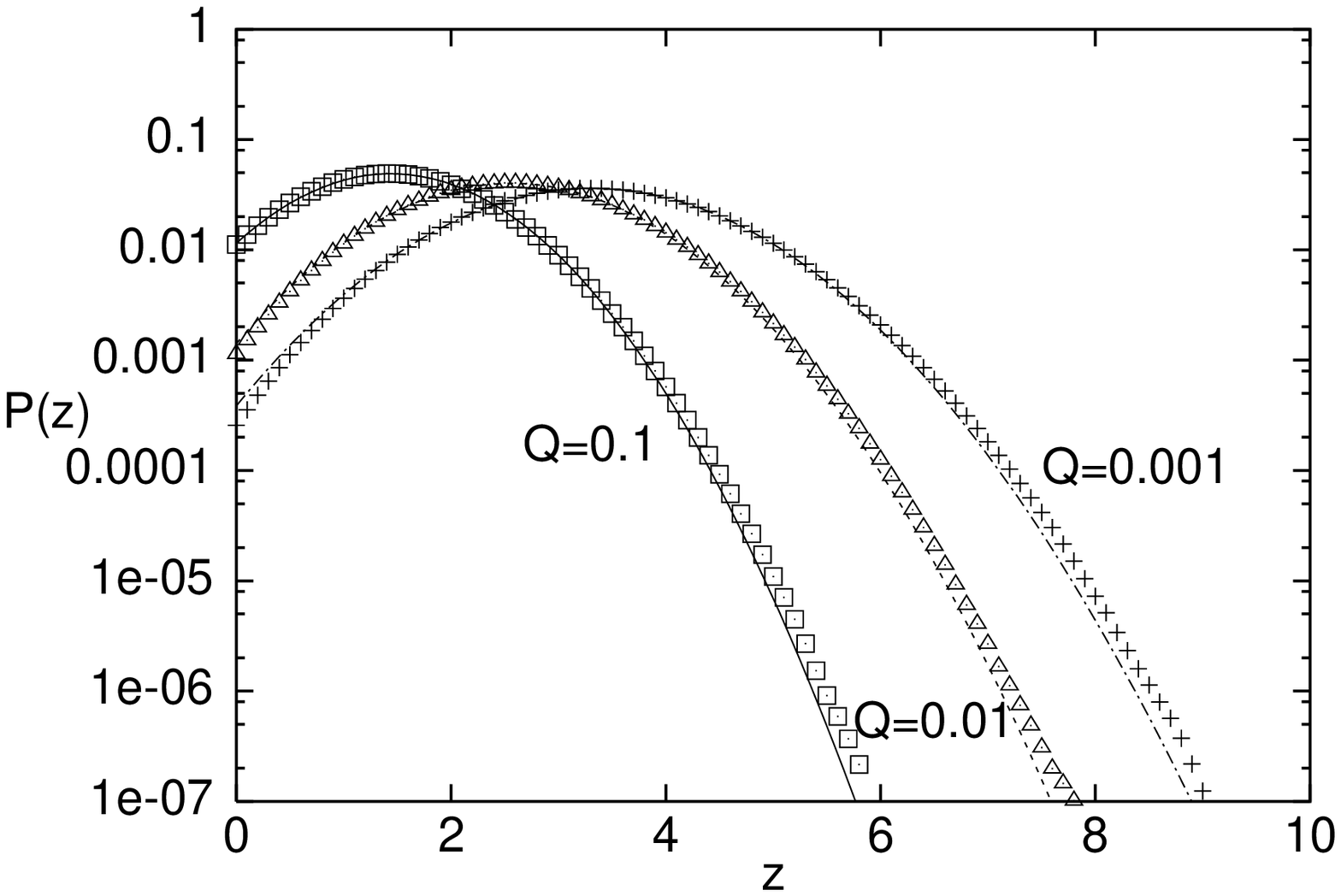}{0.45\textwidth}{Probability
  distribution $\cP(z)$ for the discrete Gau{\ss}ian model of a fluctuating
  surface which is pushed against a hard wall by the external pressure $Q$.
  The data points are fitted by Gau{\ss}ians.}

Consequently, the value $\nu_2=1$ of the contact exponent is confirmed by MC
simulations of the discrete Gau{\ss}ian model. Similar data have been
obtained in simulations of the solid--on--solid model as explained in the
next section.

In addition, the predictions for the global quantities $\ell$ and
$\xi_\perp$, as given by (\ref{ell_P}) and (\ref{xi_P}), have been
confirmed in our simulations. The corresponding data are not shown here
since they are similar to the corresponding data for the solid--on--solid
model shown in the next section. From the fits of the
mean distance $\ell$ to the functional form as given by (\ref{ell_P}), we
have extracted the dimensionless protrusion length $z_{pr}\equiv
l_{pr}/l_{sc} = 0.55\pm 0.02$. We have checked that this value of $z_{pr}$
is consistent with the fits of the roughness $\xi_{\perp}$ as in
(\ref{xi_P}).

\subsection{Solid--on--solid models}
\label{sec:SOS_HW}

In order to set up Monte Carlo simulations of the solid--on--solid
model for collective protrusions as defined by (\ref{ProtrusionHam}),
we use the 
dimensionless length $z \equiv l/l_{sc}$ and pressure $Q \equiv
P/P_{sc}$, where the length and the pressure scale are now given by
\begin{equation}
\label{l_sc_sos}
  l_{sc}=T/a_0\Sigma_0 \qquad \mbox{and} \qquad
  P_{sc}=a_0\Sigma_0/A_0.
\end{equation}

In terms of these rescaled quantities, the effective Hamiltonian  for
one protruding surface reads
\begin{equation}
  {\bar{\cal{H}}}\{z\} =
  \sum_{\langle ij \rangle} |z_i-z_j|/n_{nn} + \sum_i (Q z_i + U_{hw}(z_i)),
\end{equation}
where $n_{nn}$ is the number of nearest neighbours of each molecule as
before.
The generalization for two protruding surfaces is obvious and leads to
\begin{eqnarray}
\label{Hsos2}
  &&{\bar{\cal{H}}}\{z_1,z_2\} =
  \sum_{\langle ij \rangle} \frac{1}{n_{nn}}
  \left(|z_{1,i}-z_{1,j}|+|z_{2,i}-z_{2,j}|\right) \nonumber \\
  && \qquad +
  \sum_i Q\,(z_{1,i}-z_{2,i}) + U_{hw}(z_{1,i}-z_{2,i}),
\end{eqnarray}
which, after introducing the difference coordinate $z \equiv z_1-z_2$ and
the center--of--mass coordinate $\bar z \equiv z_1+z_2$, takes the form
\begin{eqnarray}
  {\bar{\cal{H}}} \{\bar z,z\} &= &
  \sum_{\langle ij \rangle} \frac{1}{n_{nn}}\,
  \mbox{Max}(|\bar z_i-\bar z_j|,|z_i-z_j|) \nonumber\\
  && + \sum_i Q\,z_i + U_{hw}(z_i).
\end{eqnarray}
Note that in the solid--on--solid (SOS) model, the center--of--mass
coordinate and the difference coordinate do not decouple, in contrast to the
Gau{\ss}ian model. By this decoupling, the Gau{\ss}ian model with two
surfaces can be exactly mapped to a model for two independent surfaces, one
of which is free (the center--of--mass coordinate), the other one being
subject to the external potential (the difference coordinate).
We have employed Monte Carlo simulations in order to check whether
the critical behavior of global quantities in the SOS
model for two interfaces as given by (\ref{Hsos2}), is nevertheless
consistent with the behavior obtained from the Gau{\ss}ian model.

\pic{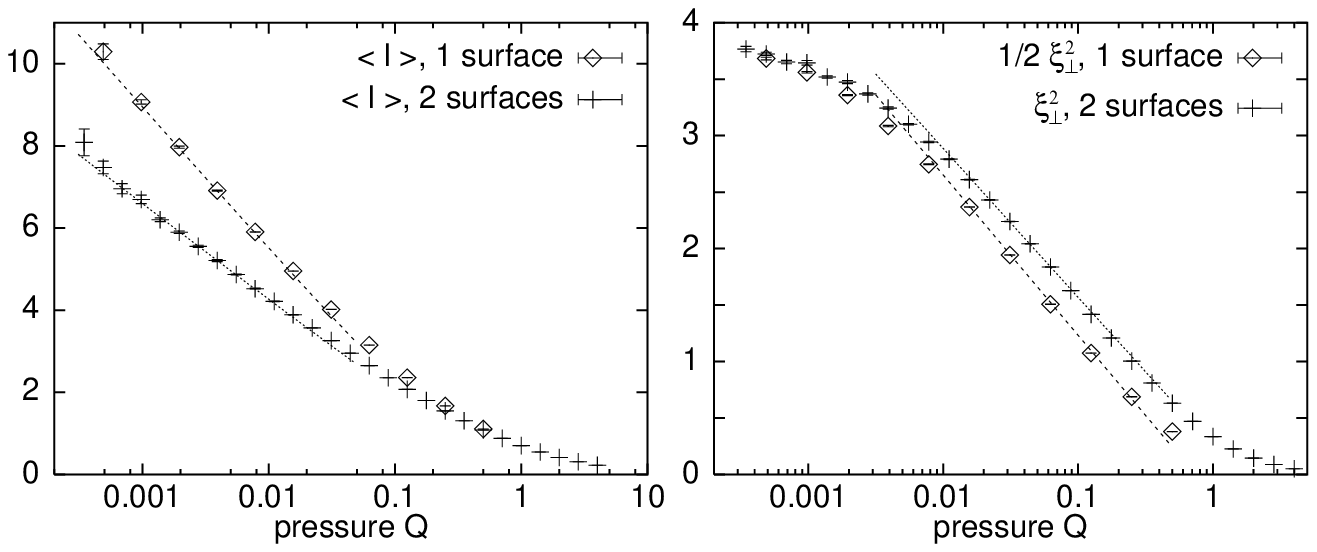}{0.495\textwidth}{MC data for the SOS model. Mean distance
  $\ell$ ({\em left}) and roughness $\xi_\bot^2$ ({\em right})
  versus pressure $Q$, both in the case of one and of two fluctuating
  surfaces. The mean distance is fitted by $\ell\simeq
  l_{pr}(\ln(Q_{hw}/Q) - 1/4 \ln\ln(Q_{hw}/Q))$, see
  Eq.~(\protect\ref{ell_P}), with $l_{pr}=1.57$ and $Q_{hw}=0.47$ for 
  one surface and $l_{pr}=1.06$, $Q_{hw}=0.81$ for two surfaces. The
  values for the roughness are fitted by
  $\xi_\perp^2\simeq l_{pr}^2/2\,\ln(Q_{hw\perp}/Q)$, see
  Eq.~(\protect\ref{xi_P}), with $Q_{hw\perp}=0.74$ for one 
  surface and $Q_{hw\perp}=1.55$ for two surfaces.}

This assumption is indeed confirmed by the results from MC
simulations, as 
shown in Fig.~\ref{hw2}. There, the mean distance
$\ell$ and the roughness $\xi_\bot$ are plotted versus the pressure
$Q$, both in the case of one and of two fluctuating surfaces.
The length scale $l_{pr(2)}$ in the latter case relates to the
protrusion length $l_{pr}$ as defined in (\ref{DefLpr}) in the
following way, using the terminology of the membrane model: Consider
the surface tension $\Sigma$ of a single membrane surface in the
Gau{\ss}ian model. First,  in the case of two identical surfaces, the
reduced surface tension $\Sigma_r$ of the distance coordinate
$l_1-l_2$ is given by $\Sigma/2$. Second, the microscopic tension
$\Sigma_0$ has 
to be multiplied by 2 since two adjacent molecules in the lipid
bilayer should move coherently in order to prevent the generation of
energetically unfavorable cavities between them. With
$\Sigma\sim\Sigma_0^2$ as in (\ref{EffectiveTension}), these two
effects lead to
$l_{pr(2)}=l_{pr}/\sqrt2$.

The actual values for $z_{pr}$, as extracted from the simulations
results in the same way as described in the preceding chapter on the
Gau{\ss}ian model, are $z_{pr}= 1.57\pm 0.03$ for one surface and
$z_{pr(2)}\simeq 1.07\pm 0.02$ for two surfaces. The corresponding
fits are shown in Fig.~\ref{hw2}. 

Let us shortly focus on the high pressure regime with $Q\gtrsim
1$. In this regime, the dimensionless external pressure $Q$ is larger
than the microscopic surface tension (which has been scaled to 1), so
the molecules move almost independently. Hence neglecting the elastic
term in the Hamiltonian, the partition sum in this regime is
asymptotically given by
\begin{equation}
  \Xi(Q) \approx \prod_{i=1}^N \int_0^{\infty} e^{-Qz_i} dz_i = \frac
  1{Q^N}, 
\end{equation}
which leads to the mean separation
\begin{equation}
  \langle z\rangle  = \frac 1{N}\sum_i \langle z_i\rangle  =
  -\frac 1{N}\frac{\partial}{\partial Q} \ln \Xi(Q)= \frac1{Q} \; .
\end{equation}
In the same way, one obtains the surface roughness
$\xi_{\bot}\sim 1/Q$. These results are displayed in
Fig.~\ref{highqlog}. 

\pic{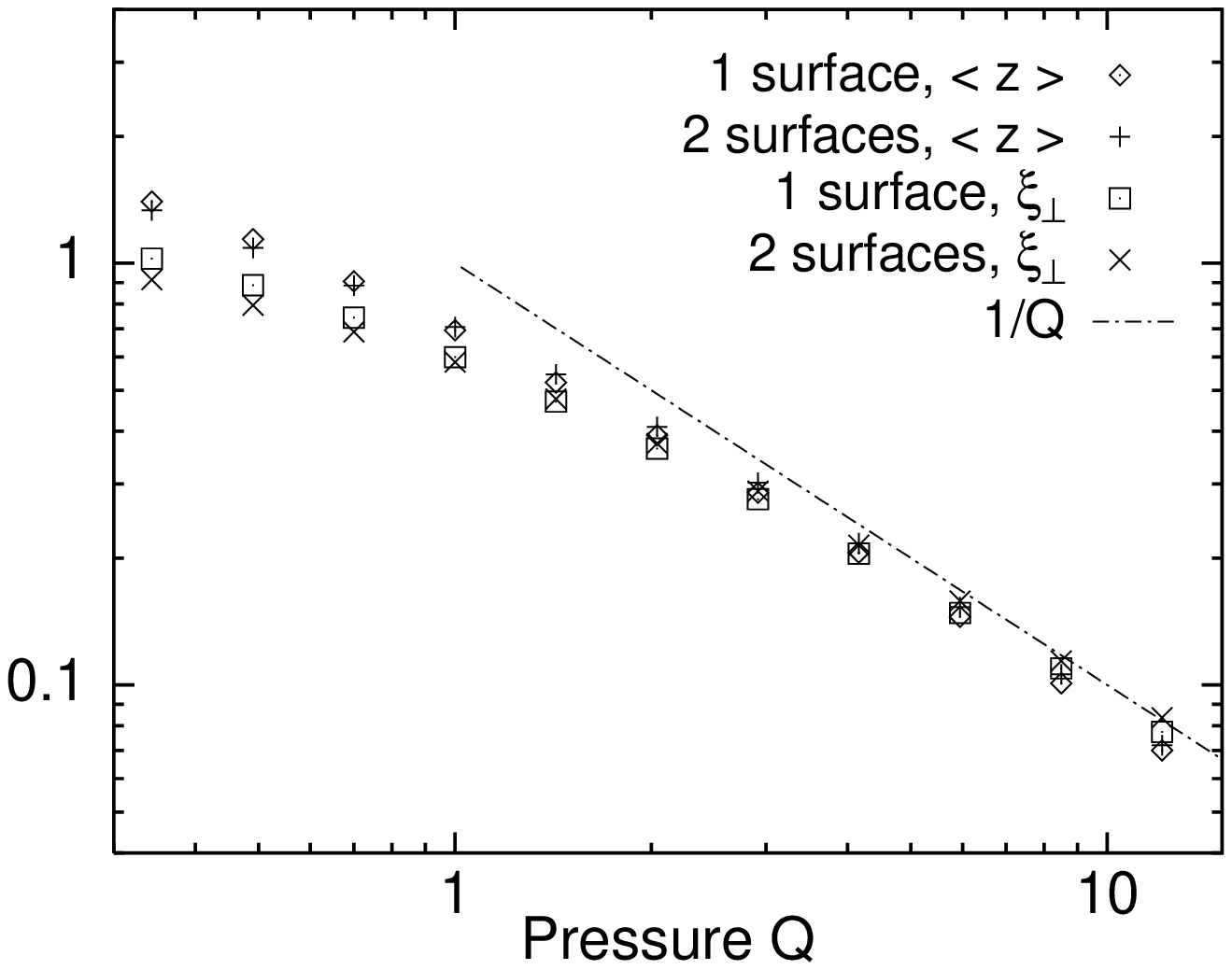}{0.4\textwidth}{
  Mean distance $\langle z \rangle$ and roughness $\xi_\bot$ versus
  pressure $Q$ in the high pressure regime for the SOS-model. The MC
  data were obtained for a pure hard wall potential and a triangular
  lattice with linear dimension $L_\parallel=32$.}

Now, let us inspect the probability distribution $\cP(l)$, as defined
in (\ref{PDeltal}), in the SOS model.
Except for its tails far away from the hard wall, $\cP(l)$  is rather
well fitted by a Gau{\ss}ian distribution of the form
\begin{equation}
 \begin{array}{llrcl} \D
  \cP_{\rm fit}(l) & =
   \D\frac{\alpha}{\sqrt{2\pi}\xis}
      \exp\left({-\D\frac{(l-\ell)^2}{2\xis^2}}\right)
        & \mbox{ for} & l>0 \\
  & =   \quad 0 \quad & \mbox{ for} &  l<0. \rule{0cm}{3ex}
    \end{array} 
  \label{gaussvert}
\end{equation}

\pic{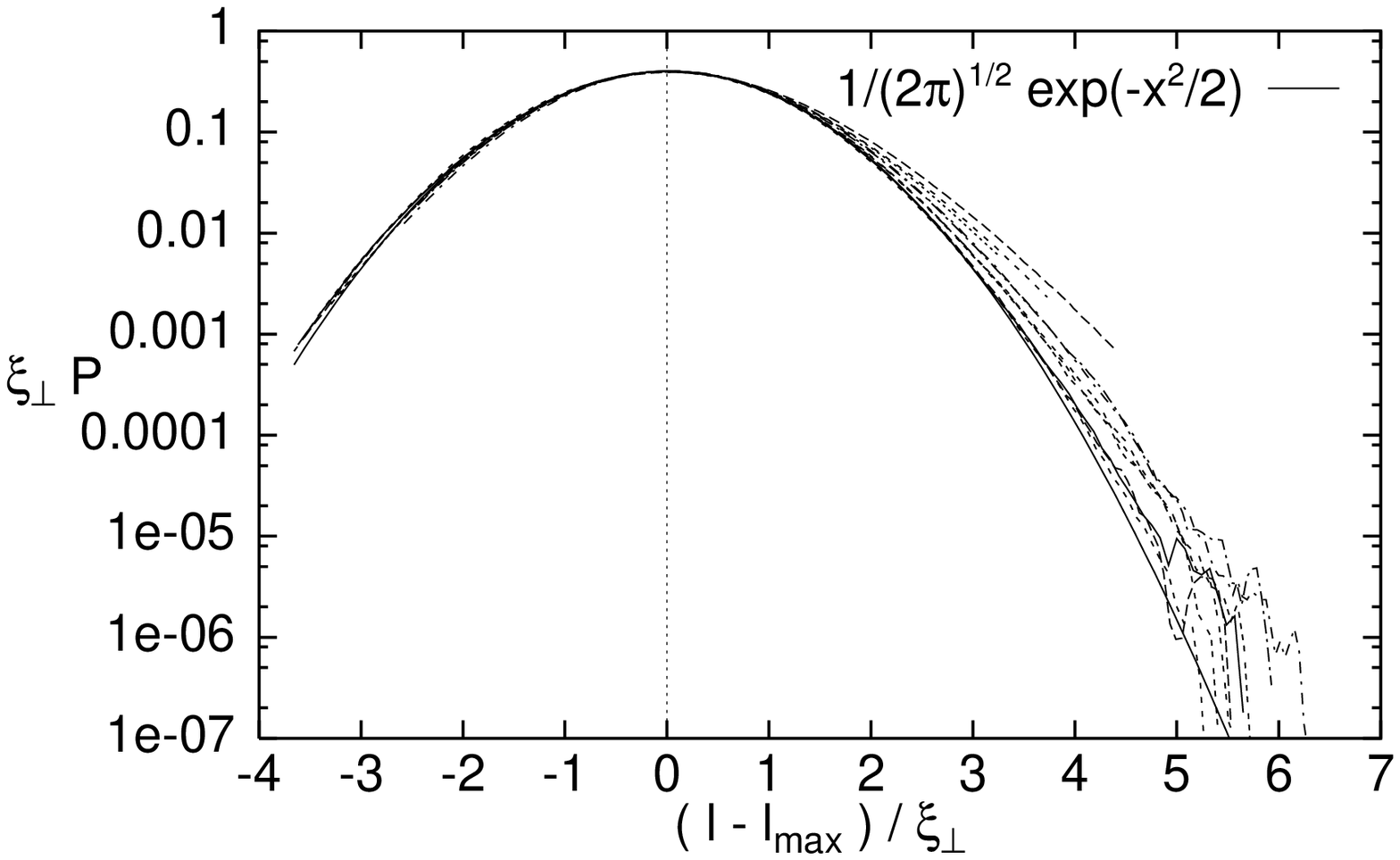}{0.4\textwidth}{Data collapse $\xi_\bot
  \cP((l-l_{max})/\xi_\bot)$ for two surfaces in the SOS model, for
  11 different values of the external pressure in the range $Q=5\times
  10^{-5} - 7\times 10^{-2}$. $l_{max}$ is the most probable distance
  between the two 
  surfaces. The topmost curve, which most strongly deviates from the
  Gau{\ss}ian, corresponds to the highest pressure value. 
}

This is shown in Fig.~\ref{pl_kollaps_2memb_hw}, where the rescaled
functions $\xi_\bot \cP((l-l_{max})/\xi_\bot)$ collapse almost onto a
single curve which coincides well with the Gau{\ss}ian distribution
$\cP_{\rm fit}(l)$.  For distances $l\lesssim\ell$, i.e., on the side
facing the hard wall, $\cP(l)$ does indeed
not deviate substantially from the Gau{\ss}ian. Hence, our argument
leading to the prediction $\nu_2=1$ for the contact exponent as in
(\ref{P_2_hw}) also applies to the SOS
model. The scaling behavior for the probability for local pair
contacts $\cP_{2b}$ is fully confirmed by numerical results which have
been obtained over a pressure range of five orders of magnitude,
see Fig.~\ref{p0_2memb}.

\pic{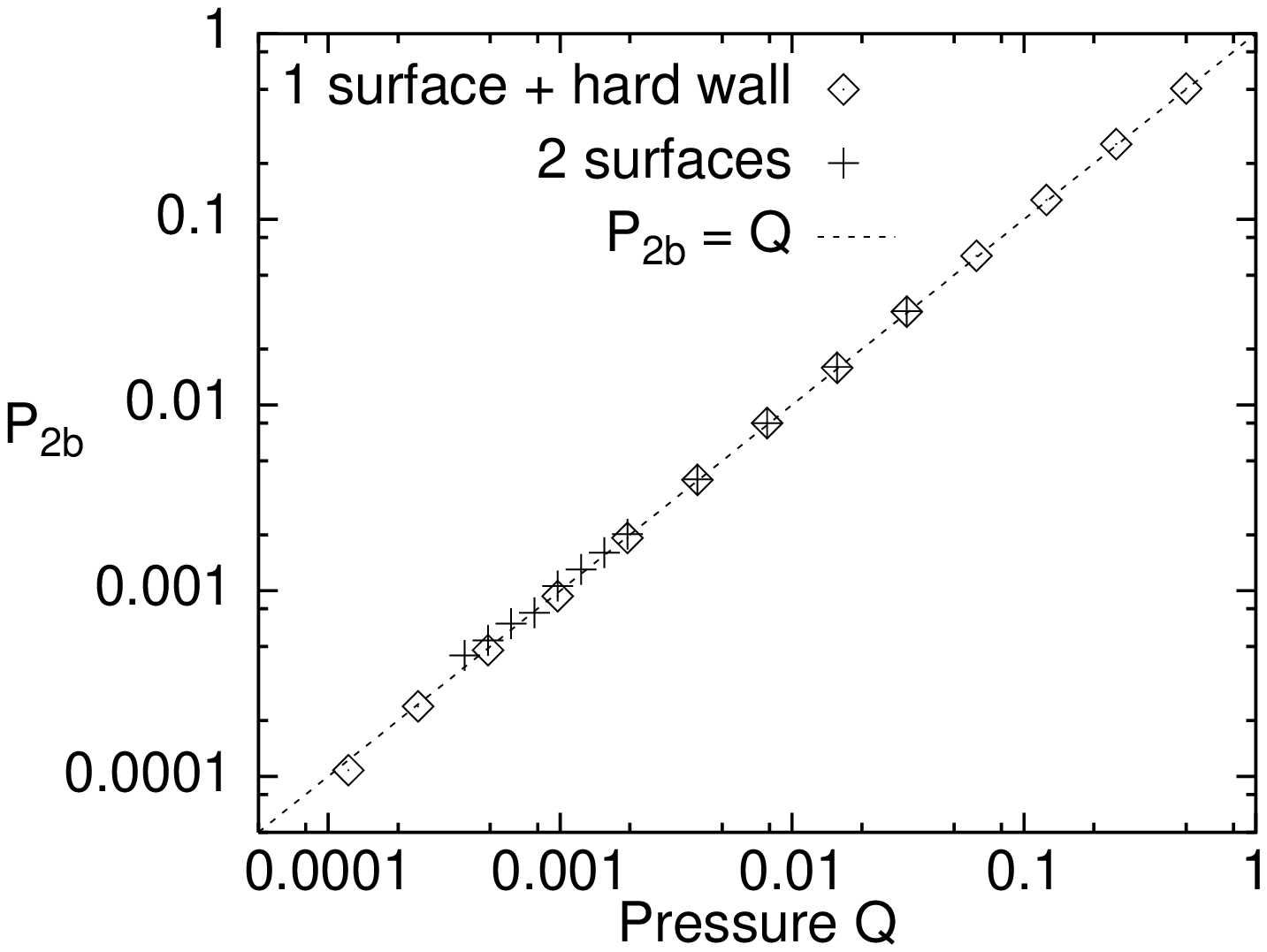}{0.35\textwidth}{Contact probability
  $\cP_{2b}$ versus pressure $Q$ for the SOS-model. The dotted
  line corresponds to the value $\nu_2=1$ for the contact exponent.
  It turns out that $\cP_{2b}=\alpha Q$ with $\alpha\simeq 1$.
}

For $l>\ell$, on the other hand, substantial deviations from
the Gau{\ss}ian behavior are observed since the histograms have an
exponential tail; see also Fig.~\ref{hist_1}. This tail results from
isolated 
protrusions, which can be understood as follows. The energy of an
isolated protrusion of height $l$ is
given  by $E(l)=\sum_{n.n.}|l_i-l|/(n_{nn}l_{sc})$, which, if all
neighbouring sites $i$ have heights $l_i<l$, simplifies to $E(l)=|\bar
l_{nn}-l|/l_{sc}$,
with $\bar l_{nn}=n_{nn}^{-1}\sum_{n.n.}l_i$. Together with the
external pressure term $Q$, the probability of a single
protrusion is hence proportional to $e^{-z(1+Q)}$. The fits in
Fig.~\ref{hist_1} confirm this simple explanation.

\pic{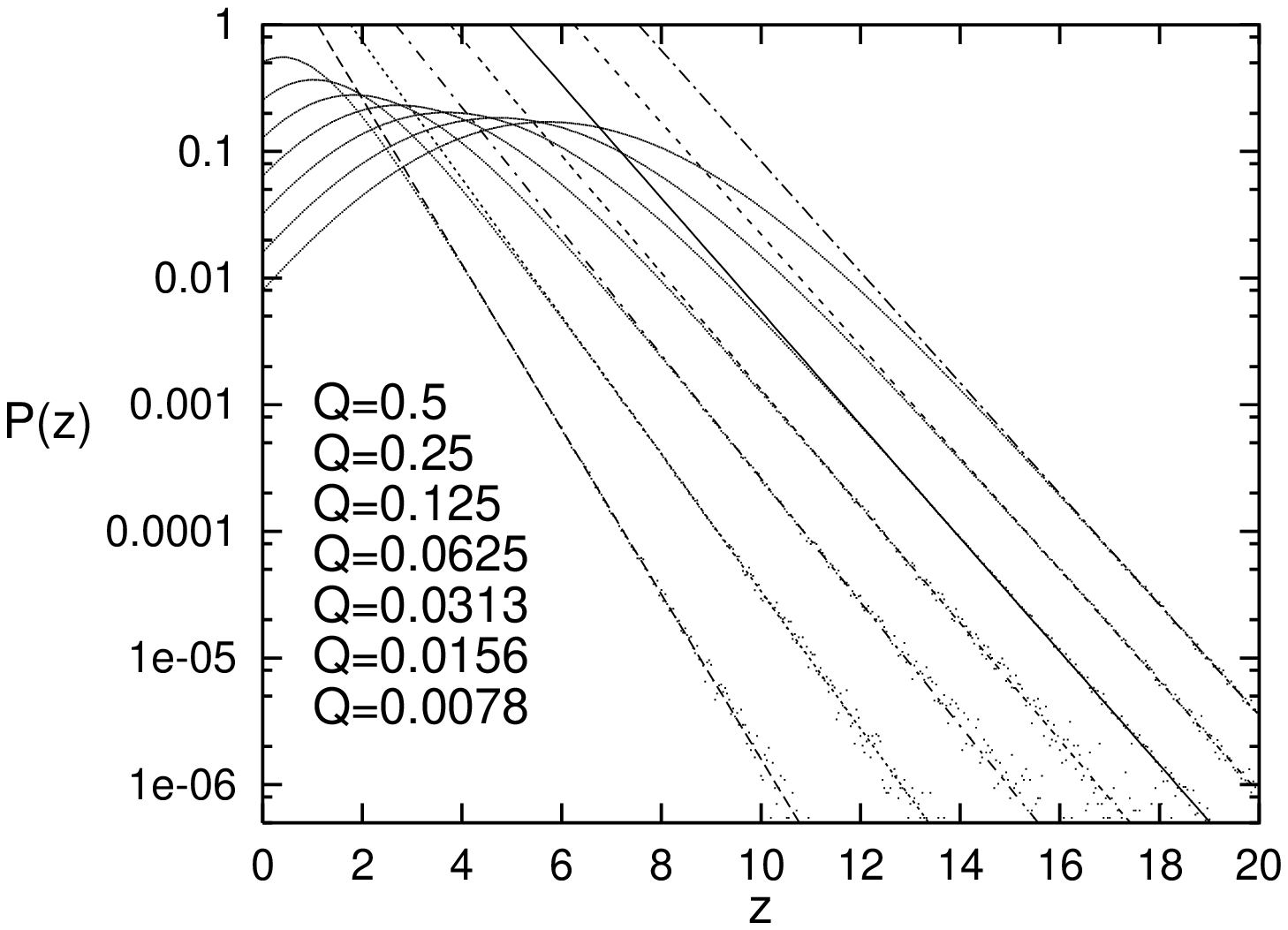}{0.45\textwidth}{Probability distribution
  $\cP(z)$ for one fluctuating surface with hard wall interaction
  within the SOS model. The 
  tails have been fitted by exponentials of the form $c\,e^{-z(1+Q)}$,
  where the amplitude $c$ represents a fit parameter.}

\section{Two surfaces with soft wall interactions}

\label{Sec:5} 

Now, we will study two surfaces which are governed by the potential
\begin{equation}
  V(l) = V_{hw} (l) + V_{hy} \exp[- l / l_{hy}] + Pl
\end{equation}
which includes the hydration interaction
$V_{hy} e^{- l / l_{hy}}$. The decay length of this soft wall is
given by the hydration length $l_{hy}$, which introduces a new length
scale into the problem.

Linear functional renormalization of the hydration interaction shows
that the interplay between the protrusion length $l_{pr}$ and the
hydration length $l_{hy}$ leads to two different regimes
\cite{lipo105+lipo108}: For $l_{pr} > 2 \,l_{hy}$ and $l_{pr} < 2
\,l_{hy}$, one has a {\em protrusion regime} and a {\em hydration
regime}, respectively. In the protrusion regime, the asymptotic
behavior is the same as in the case of a hard--wall interaction
alone. In the hydration regime, on the other hand, one finds
qualitatively new behavior.

Within the hydration regime, the relation between the pressure $P$ and
the interfacial roughness $\xi_\bot$ has again the Gau{\ss}ian form as
given by (\ref{P_xis}) but with the new pressure amplitude
$P_{2\perp}=l_{hy}\Sigma/a^2$. The relation between the pressure and
the mean distance $\ell$, on the other hand, is now given by
\begin{equation}
\label{P_lt}
P \approx P_2\, e^{-\ell/l_t}
\end{equation}
which is governed by the new length scale
\begin{equation}
\label{l_t}
l_t = l_{hy}\,\left[1+(l_{pr}/2l_{hy})^2\right].
\end{equation}
The pressure amplitude $P_2$ has the parameter dependence
\begin{equation}
  \label{P_2_amp}
  P_2 = (\Sigma/a^2)^{\rho}\;V_{hy}^{1-\rho}
        \;l_{hy}^{2\rho-1}
\end{equation}
with $\rho\equiv 1-l_{hy}/l_t$.

Note that the length scale $l_t$ as given by (\ref{l_t}) depends both
on the protrusion length $l_{pr}$ and on the hydration length
$l_{hy}$. For $l_{hy}\gg l_{pr}$, one has $l_t \approx l_{hy}$; i.e.,
in the limit in which the hydration potential has a much longer range
than the effective protrusion potential, the mean separation is solely
determined by the balance between the applied pressure and the
repulsive hydration forces.


\subsection{Gau{\ss}ian models}
\label{sec:5.1}

Let us introduce two additional dimensionless quantities for
the soft wall interaction via
\begin{equation}
\label{z_hy_U_hy}
  z_{hy} \equiv l_{hy} / l_{sc} \qquad \mbox{and} \qquad
  U_{hy} \equiv A_0 V_{hy}/T.
\end{equation}
The above results for the pressure-dependence of $\ell$ and $\xi_\bot$ are
confirmed by MC simulations of the discrete Gau{\ss}ian model,
as shown in Fig.~\ref{GaussZvsQ} and Fig.~\ref{GaussXIvsQ}.

\pic{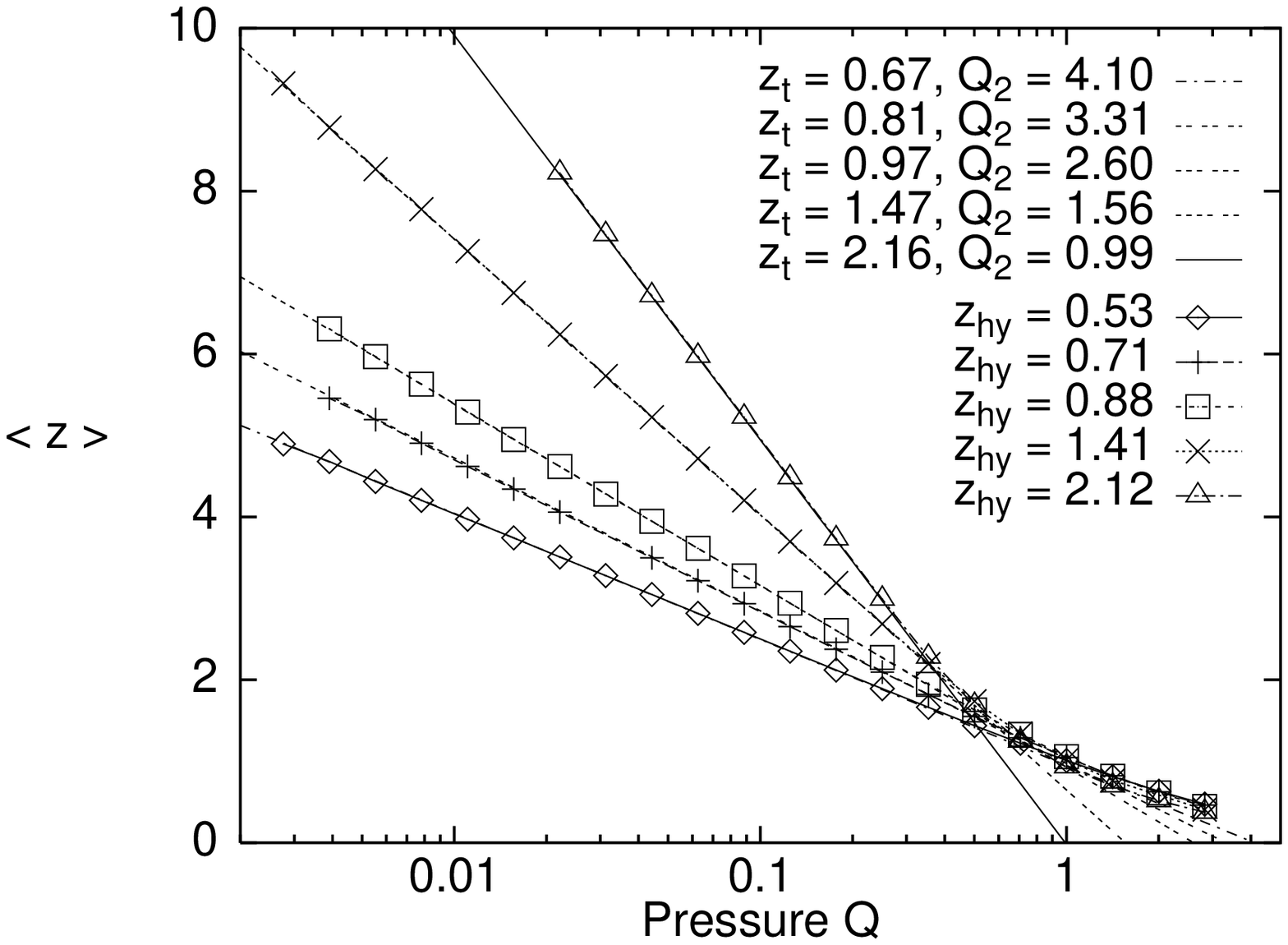}{0.4\textwidth}{Discrete Gau{\ss}ian model:
  Mean distance $\langle z \rangle$ versus pressure $Q$ for several
  hydration lengths $z_{hy}$ for one surface fluctuating against a
  soft wall, with a linear dimension $L_\parallel=64$ and
  potential strength $U_{hy}=5.7$. The data are fitted by the
  functional form $z=z_t\ln(Q_2/Q)$.}  

\pic{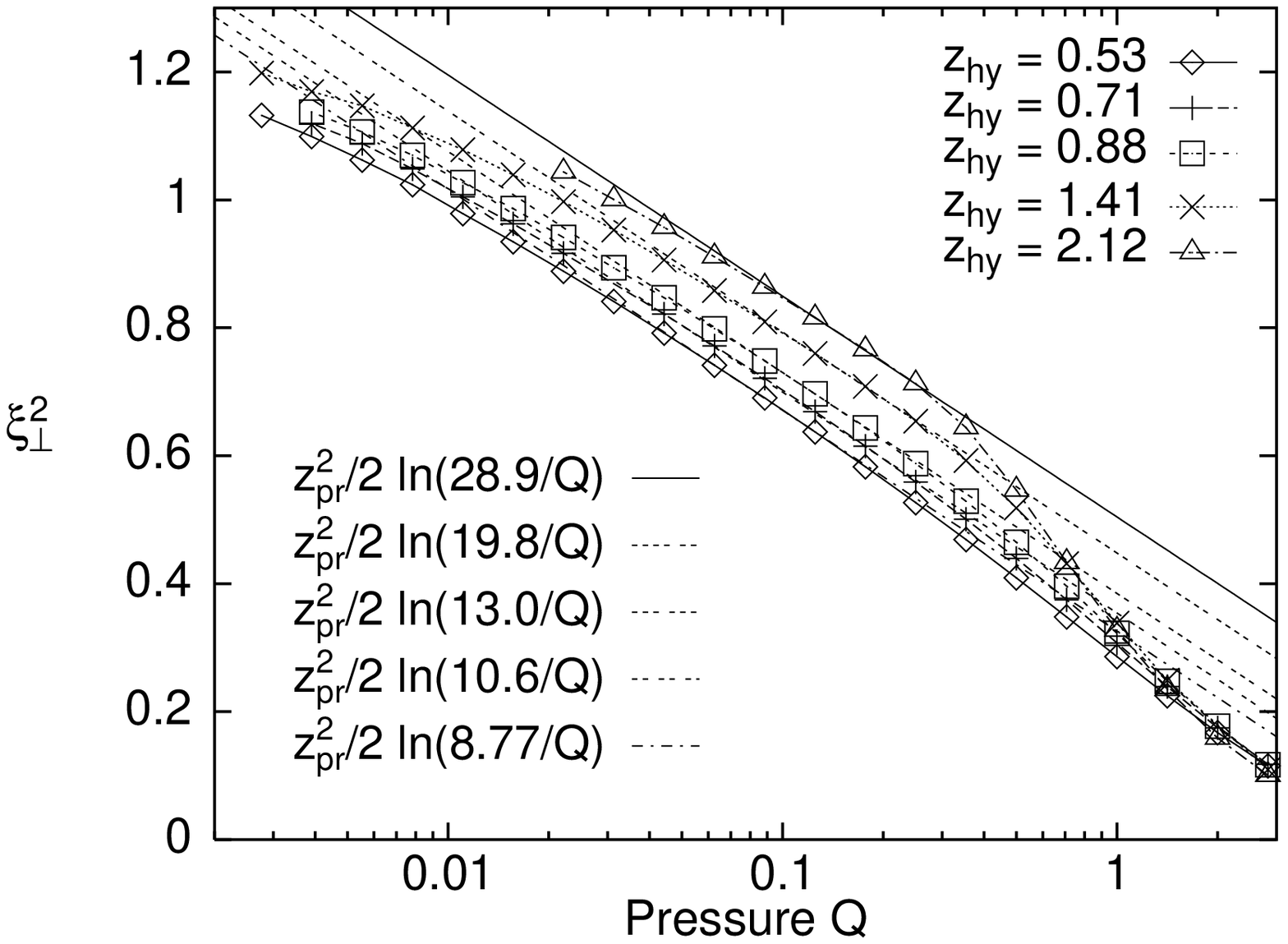}{0.4\textwidth}{Discrete Gau{\ss}ian
  model: Roughness $\xi_\bot^2$ versus pressure $Q$ from the same MC
  data as in Fig.~\protect\ref{GaussZvsQ}.
  For the fit, the value $z_{pr}^2/2=0.15$
  has been used, corresponding to $z_{pr}\simeq 0.55$. Note the finite
  size effects for $Q\lesssim 0.02$.}

As we have already shown in the preceding section, the probability
distribution $\cP(l)$ for the discrete Gau{\ss}ian model with
a hard wall interaction has no exponential tail. In the presence of
the hydration interaction, this distribution is well fitted by the
functional form
\begin{equation}
\label{fit_Pl_uhy}
  \cP_{Gau}(l) \sim \exp\left[
    -\frac{(l-\ell)^2}{2\xi_\bot^2} -
    U_{hy}e^{-l/l_{hy}} \right],
\end{equation}
where the influence of the direct hydration potential has been taken
into account in the simplest possible manner. Note that the term that
explicitely depends on the hydration potential in (\ref{fit_Pl_uhy})
is appreciable only for $l \lesssim l_{hy}$, i.e. close to the hard wall,
while $\ell$ diverges in the limit of small $P$.
Fig.~\ref{hist_gauss} shows some fits of this form. Obviously, the
tails have pronounced Gau{\ss}ian characterists, rather than
exponential tails, on {\em both} sides. 
On the side facing the wall, the fits are excellent, while they are
slightly poorer on the other side.

\pic{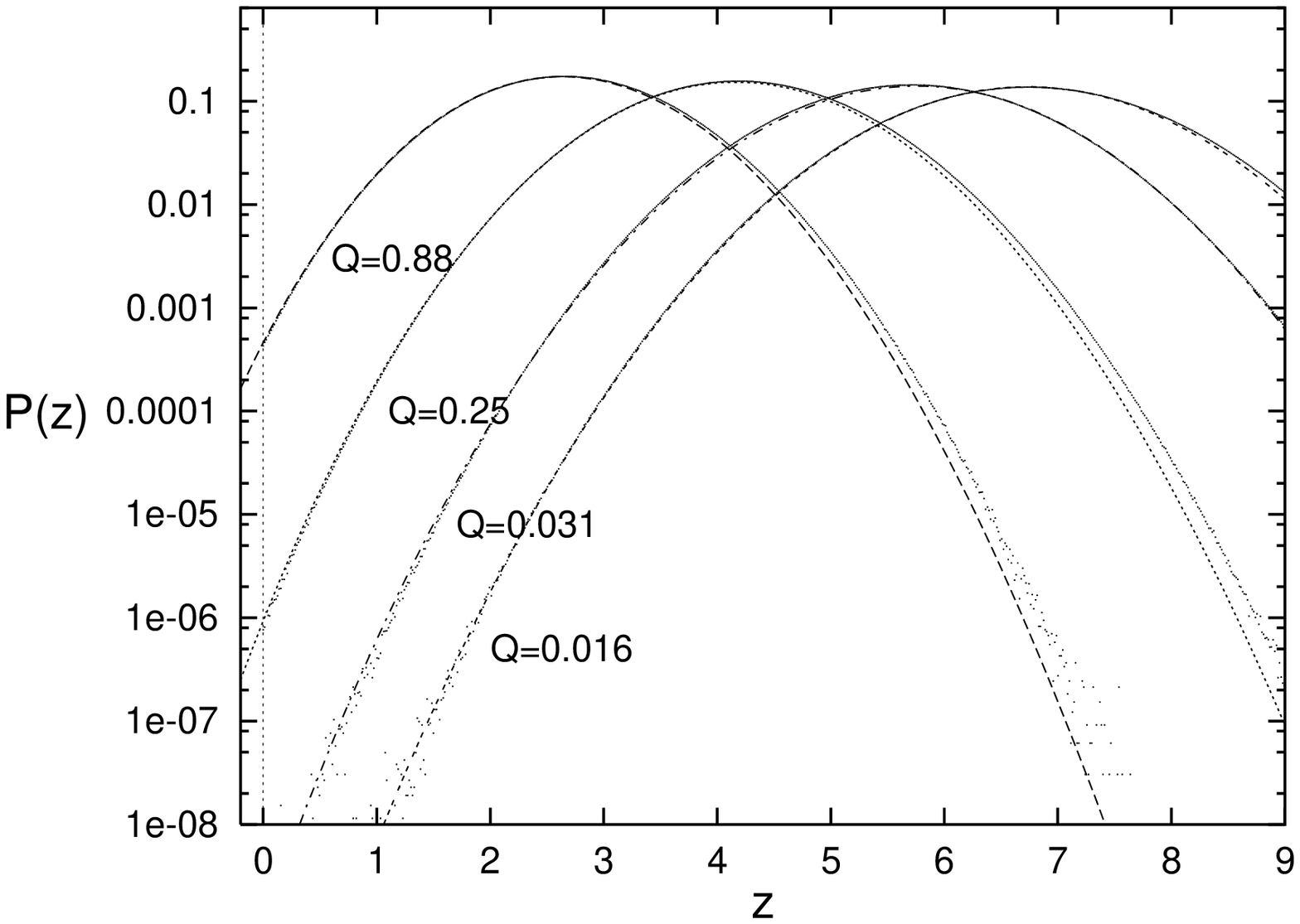}{0.495\textwidth}{Typical histograms
  for the Gau{\ss}ian model with a soft wall. Sample size 64x64,
  $U_{hy}=5.7$, $z_{hy}=1.41$. The fitting curves have the
  functional form $\exp(-(z-\langle z \rangle)^2/2\xi_\bot^2 -
  U_{hy}e^{-z/z_{hy}})$. Note the absence of exponential tails.}

An evaluation of the Gau{\ss}ian distribution $\cP_{Gau}(l)$ at the
position $l=0$ of the hard wall,
leads to the contact probability
\begin{equation}
\label{P2b_gaussian}
  \cP_{2b} \sim e^{-\ell^2/2\xis^2}
  \sim P^{(l_t/l_{pr})^2},
\end{equation}
in the limit of small pressure $P$, where the pressure dependence of $\xis$
and $\ell$ as given by (\ref{xi_P}) and (\ref{P_lt}), resp., has
been used, ignoring confluent logarithmic terms.

Thus, the contact exponent $\nu_2$ is found to be
\begin{equation}
\label{nu2_gaussian_uhy}
  \nu_2=(l_t/l_{pr})^2,
\end{equation}
where the length scale $l_t$ is given by (\ref{l_t}). Note that for
$l_{hy}=l_{pr}/2$, which defines the border line between the protrusion
and the hydration regime, one has $l_t=l_{pr}$ and hence $\nu_2=1$ as in the
protrusion regime. Therefore, the contact exponent $\nu_2$ is a continuous
function of
$\l_{hy}$.

The corresponding data from MC simulations of the discrete Gau{\ss}ian
model are shown in Fig.~\ref{GaussNu2}. An inspection of this figure reveals
considerable deviations from the predicted values as given by
(\ref{nu2_gaussian_uhy}). These deviations can be explained in the
following way.

\pic{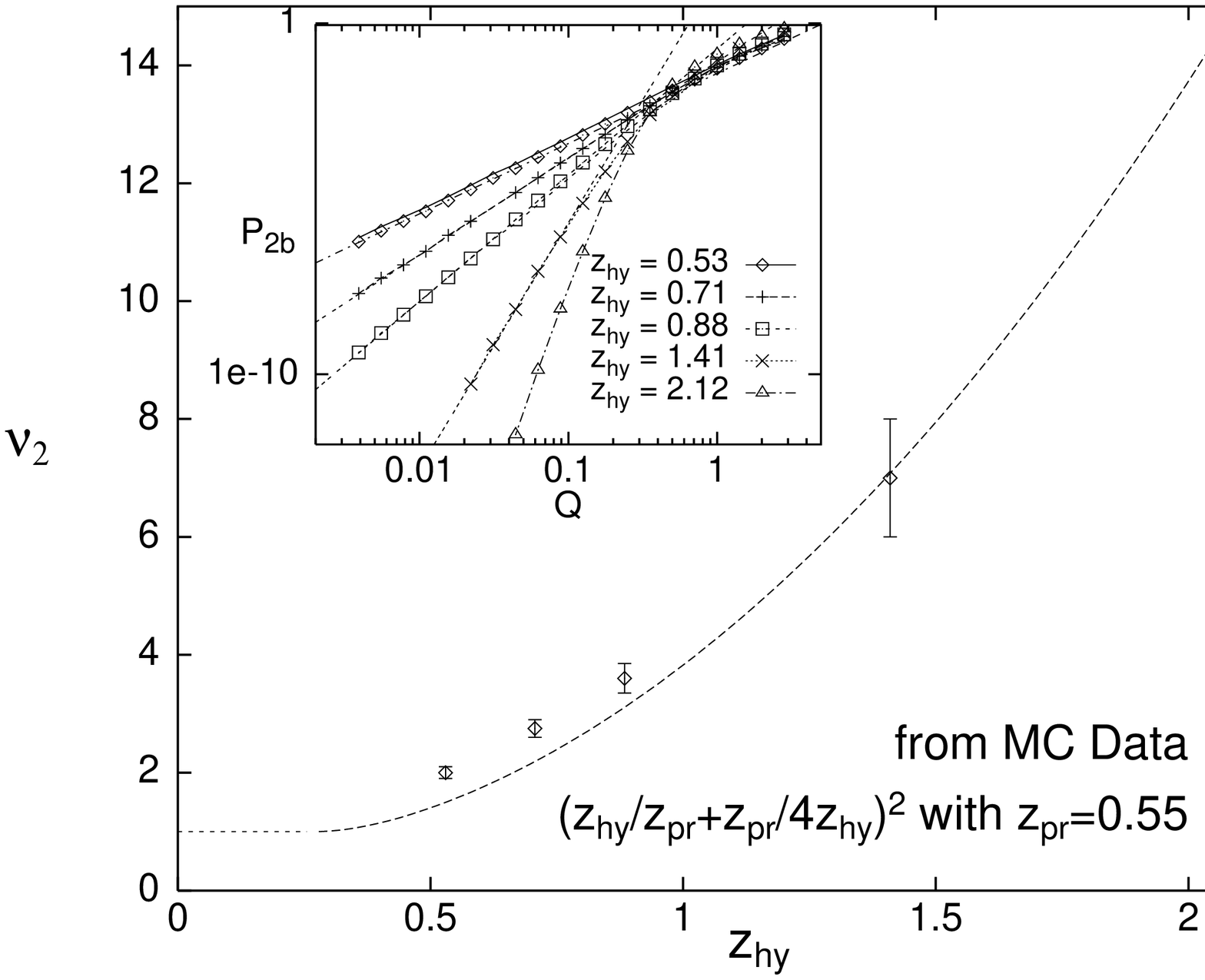}{0.495\textwidth}{Contact exponents $\nu_2$ as a
  function of the hydration length $z_{hy}$ for the discrete
  Gau{\ss}ian model. In the inset, fits to MC data for the contact
  probability $\cP_{2b}$ versus pressure $Q$ are shown.}

First of all, the accessible regime of pressure values $Q$ is quite
restricted, covering only about one order of magnitude: For large $Q>
1$, we are deep in the mean field regime where the collective
behavior is dominated by the external pressure. In the latter case, the
molecules move almost independently and the probability distibution is
$\sim Q \exp(- Qz)$. This implies that the contact probability scales
linearly with
$Q$, not depending on the details of the hydration force. For small
$Q\lesssim 0.01$, on the other hand, the finite size of the sample effects
the interface roughness
$\xi_{\perp}^2$, see~Fig.~\ref{GaussXIvsQ}. This has a large effect on the
histogram, making it 'narrower' than it would be for larger systems, while
the mean distance $\ell$ is essentially unaffected by the finite size. This
lowers the observed contact probability $\cP_{2b}\sim\cP(0)$. For
this reason, we expect to extract a contact exponent from the data
which is larger than the true asymptotic value.
This behavior is expected for all values of $z_{hy}$, hence all data
will tend to lie {\em above} the theoretical line.

Secondly, the asymptotic form for $\cP_{2b}$
as given by (\ref{P2b_gaussian}) is difficult to extract from
intermediate values of $Q$ because of large corrections to the leading
scaling behavior which arise as long as the pressure amplitudes $Q_2$ and
$Q_{2\perp}$ take different values. Indeed, one has $\cP_{2b} \sim \exp [ -
\ell^{2} / 2 \xi_{perp}^{2} ]$ with the asymptotic behavior
\begin{eqnarray}
\label{P2b_hy_intermediate}
  \frac{\ell^2}{2\xi_{\perp}^2} &\approx&
  \frac{l_t^2}{l_{pr}^2}\;
  \frac{\ln^2(Q_2/Q)}{\ln(Q_{2\perp}/Q)} \\
  &=&
  \frac{l_t^2}{l_{pr}^2}\,\left[  
    \ln\left( 
      \frac{Q_2^2}{Q_{2\perp} Q}
    \right)
  - \frac{ \ln^2(Q_2/Q_{2\perp}) }{\ln Q} +
  {\cal O}\left( \frac{1}{\ln^2 Q} \right) \right]. \nonumber
\end{eqnarray}
The leading order correction vanishes only for $Q_2=Q_{2\perp}$.
Using the values for $Q_2$ and $Q_{2\perp}$ as measured in the MC
simulations of the discrete Gau{\ss}ian model, we find that these corrections
affect our data both for
$z_{hy}=1.4$ and for $z_{hy}=2.1$.

The measured contact exponents are however consistent with
the actual values of $\exp[-\ell^2/2\xi_{\perp)}^2]$
as determined from the measured values of $\ell$ and $\xi_\perp$. In
order to show this, we have plotted
$\exp(-l_t^2\ln^2(Q_2/Q)\;/\;l_{pr}^2 \ln(Q_{2\perp}/Q))$,
corresponding to the first equality in (\ref{P2b_hy_intermediate}), as
a function of $Q$, see Fig.~\ref{CheckAsympt} ({\em full lines}). The
values for $Q_2$ and $Q_{2\perp}$ were taken from the MC data. For
comparison, we have 
included the expected asymptotic scaling function $\cP_{2b}\sim
Q^{z_t^2/z_{pr}^2}$, with the 
contact exponents $\nu_2=z_t^2/z_{pr}^2=7.14$ for $z_{hy}=1.4$ and
$\nu_2$=15.4 for $z_{hy}$=2.1 ({\em dashed lines}).
Finally, for the latter value $z_{hy}=2.1$, the apparent scaling in
the intermediate regime $0.03\lesssim Q \lesssim 0.2$
with the exponent $\nu_2=13$ (as determined from MC data, see
Fig.~\protect\ref{GaussNu2}) is
displayed ({\em dashed-dotted line}). Obviously, in the pressure
regime that was accessible to MC 
simulations, this effective scaling exponent coincides well with the
theoretical prediction. Note that this effect leads to data that are
{\em lower} than the theoretical value for large values of $z_{hy}$.

Together, these two effects explain the deviations of the measured
values for $\nu_2$ from the theoretical prediction
$\nu_2=(l_t/l_{pr})^2$.

\pic{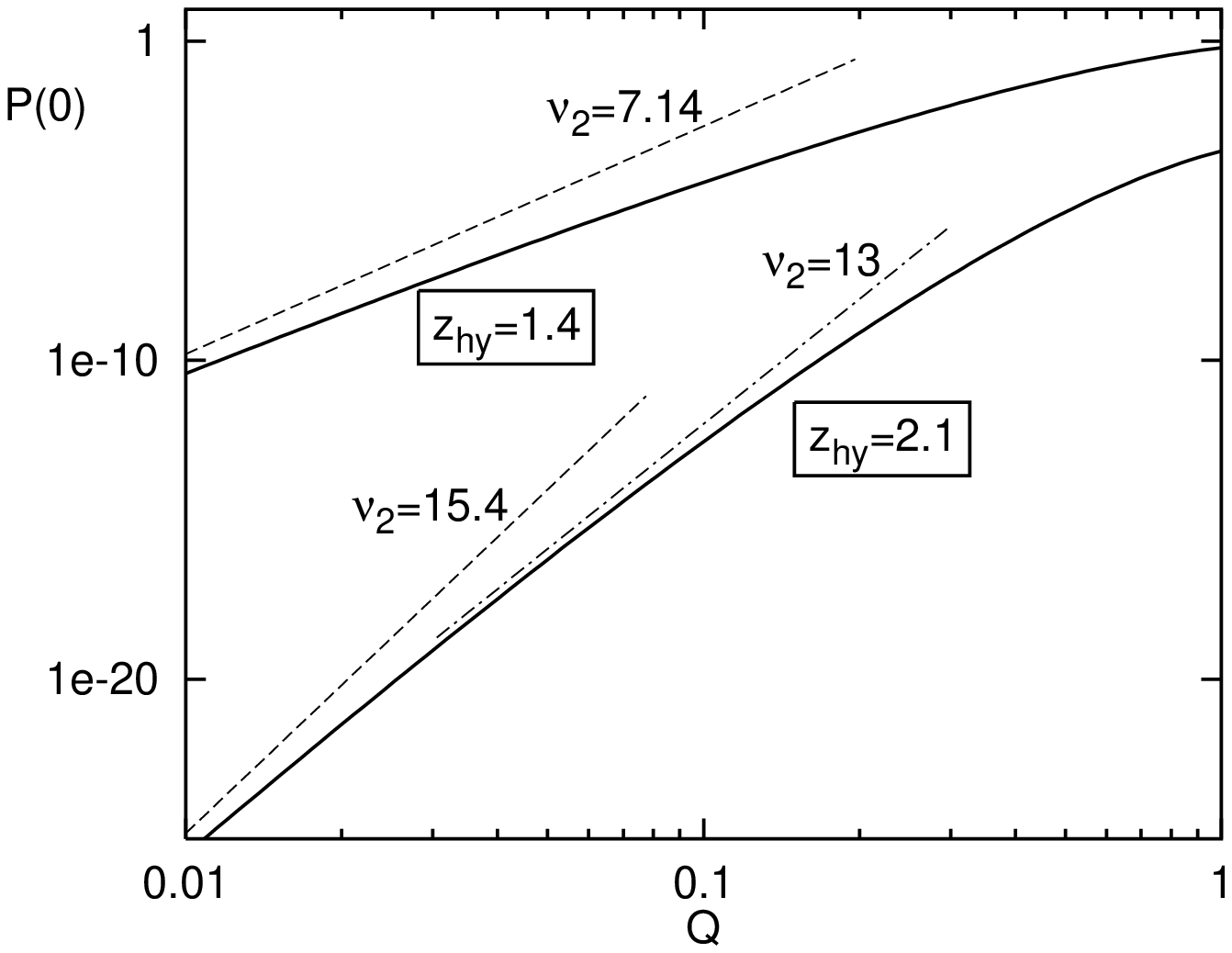}{0.48\textwidth}{Contact exponents
  for the discrete Gau{\ss}ian model with soft wall interaction. The
  upper curve corresponds to $z_{hy}=1.4$, the lower one to
  $z_{hy}=2.1$. For a detailed discussion of this figure, see section
  \protect\ref{sec:5.1}.
}

\subsection{Solid--on--solid models}

The SOS model with soft wall interactions has been expressed in terms
of the dimensionless variables (\ref{z_hy_U_hy}) and
(\ref{l_sc_sos}). Using this  dimensionless form of the model, we have
first studied the critical 
behavior of the mean separation $\ell$ and the roughness
$\xi_{\perp}$. The corresponding MC data for $z_{hy}>z_{pr}/2$
are displayed in Fig.~\ref{zt_vs_x} and Fig.~\ref{qamp_hy}. In these
figures, we compare the critical behavior of (i) one fluctuating surface
interacting with a hard wall and of (ii) two fluctuating
surfaces. Inspection of these figures shows that both cases exhibit
analogous critical behavior.  We have also studied the critical
behavior of $\ell$ and $\xi_{\perp}$ for smaller values of
$z_{hy}$. All of these data are again consistent with two distinct
scaling regimes: a hydration regime for $z_{hy} > z_{pr}/2$ and a
protrusion regime for $z_{hy} < z_{pr} /2$.

\pic{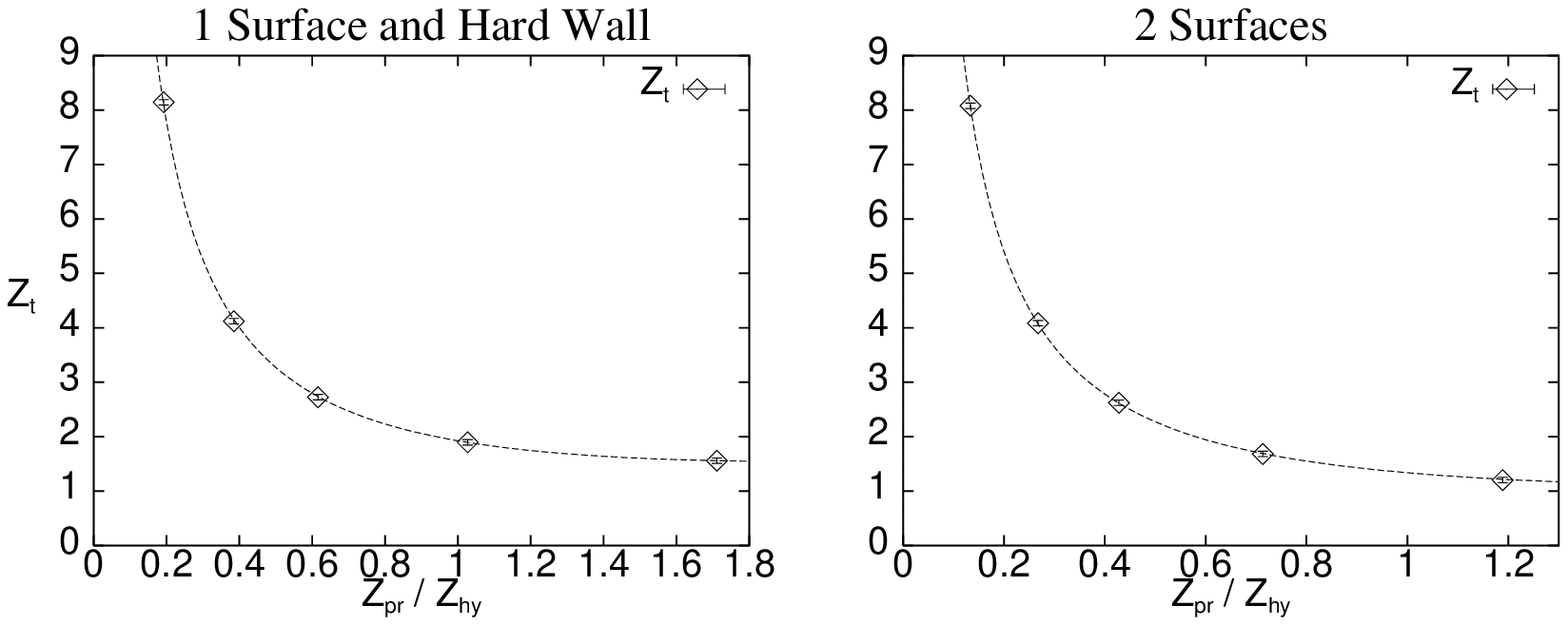}{0.495\textwidth}{Hydration regime in the SOS model:
  The length scale $z_t$ is plotted versus $z_{pr}/z_{hy}$, for one
  and for two fluctuating surfaces. The broken curve corresponds to
  the prediction of functional renormalization as given by 
  (\protect\ref{l_t}), $z_t=z_{pr}(z_{hy}/z_{pr}+z_{pr}/4z_{hy})$,
  with $z_{pr}=1.54$ for one surface and $z_{pr}=1.07$ for two
  surfaces.} 

\pic{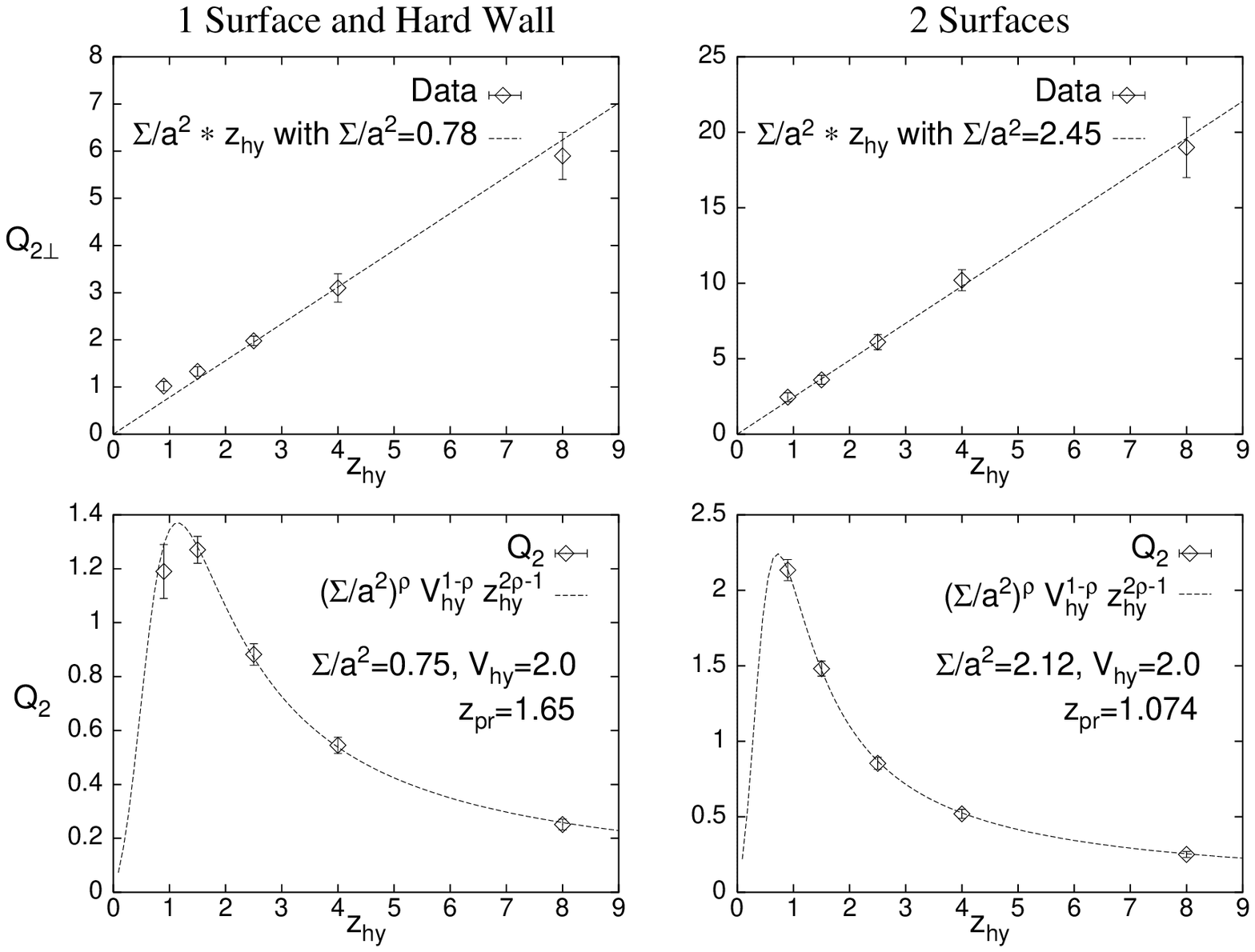}{0.495\textwidth}{The pressure amplitudes
  $Q_2$ and $Q_{2\bot}$ versus hydration length $z_{hy}$, for the SOS
  model of one and two fluctuating surfaces, respectively.  For
  comparison, the formulae from Eq.~(\protect\ref{xi_P}) are plotted.}

Next, let us consider the contact probability $\cP_{2b}$. In the
protrusion regime, the mean separation between the membranes and their
width are determined solely by the protrusion length $l_{pr}$, so the
hydration interaction should not alter the pressure dependence of
$\cP_{2b}$. Thus, one expects again $\nu_2=1$. This is confirmed by
numerical results that give the values $\nu_{2}\simeq 1.015$ and
$1.025$ for $z_{hy}=0.3$ and $0.5$, respectively, which coincides with
the expected value within the numerical precision, see
Fig.~\ref{kleineZhy_p2b}.

\pic{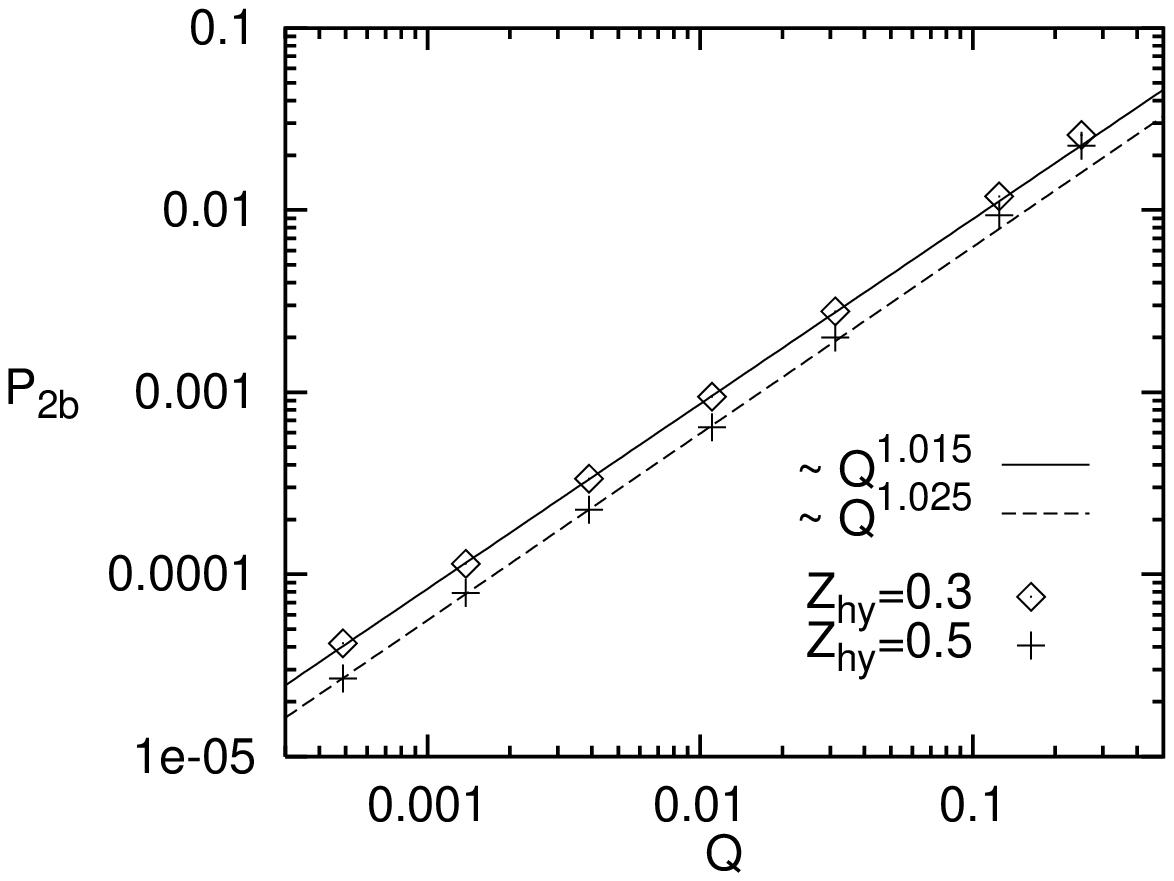}{0.35\textwidth}{Protrusion regime in the SOS
  model: Contact probability $\cP_{2b}$ versus pressure $Q$ with
  $z_{hy}<z_{pr}/2$. The data confirm the universal value $\nu_2=1$
  for the contact exponent in this case.}

The situation changes when one enters the hydration regime with
$z_{hy}>z_{pr}/2$. Now, the mean distance is proportional to $l_{t}$,
while the roughness is still determined by $l_{pr}$.
In the preceding chapter, we have found the
value $\nu_{2}=(l_t/l_{pr})^2$
for the hydration regime of the {\em Gau{\ss}ian} model.
However, we will now argue that, in the hydration regime of the {\em
  SOS} model, the contact probability $\cP_{2b}$ is strongly affected
by exponential tails of the probability distribution arising from
molecular protrusions. 

In the protrusion regime of the SOS model, the probability distribution
$\cP(l)$ has a pronounced exponential tail
on the side of the interface which does not face the wall, while on the side
facing the wall, it is essentially Gau{\ss}ian shaped. This is the
same behavior as observed in the case of pure hard wall interactions,
as discussed in section \ref{sec:SOS_HW}.
Now, in the hydration regime of the SOS model, exponential tails are
found on {\em both} sides of the histogram, as shown in
Fig.~\ref{hist_sos_hy}. 
This observation inspires a theoretical prediction for the contact
exponent that yields a much better description of the data, based
on the assumption that these tails are the result of isolated
protrusions which can be considered as 'rare' events.

\pic{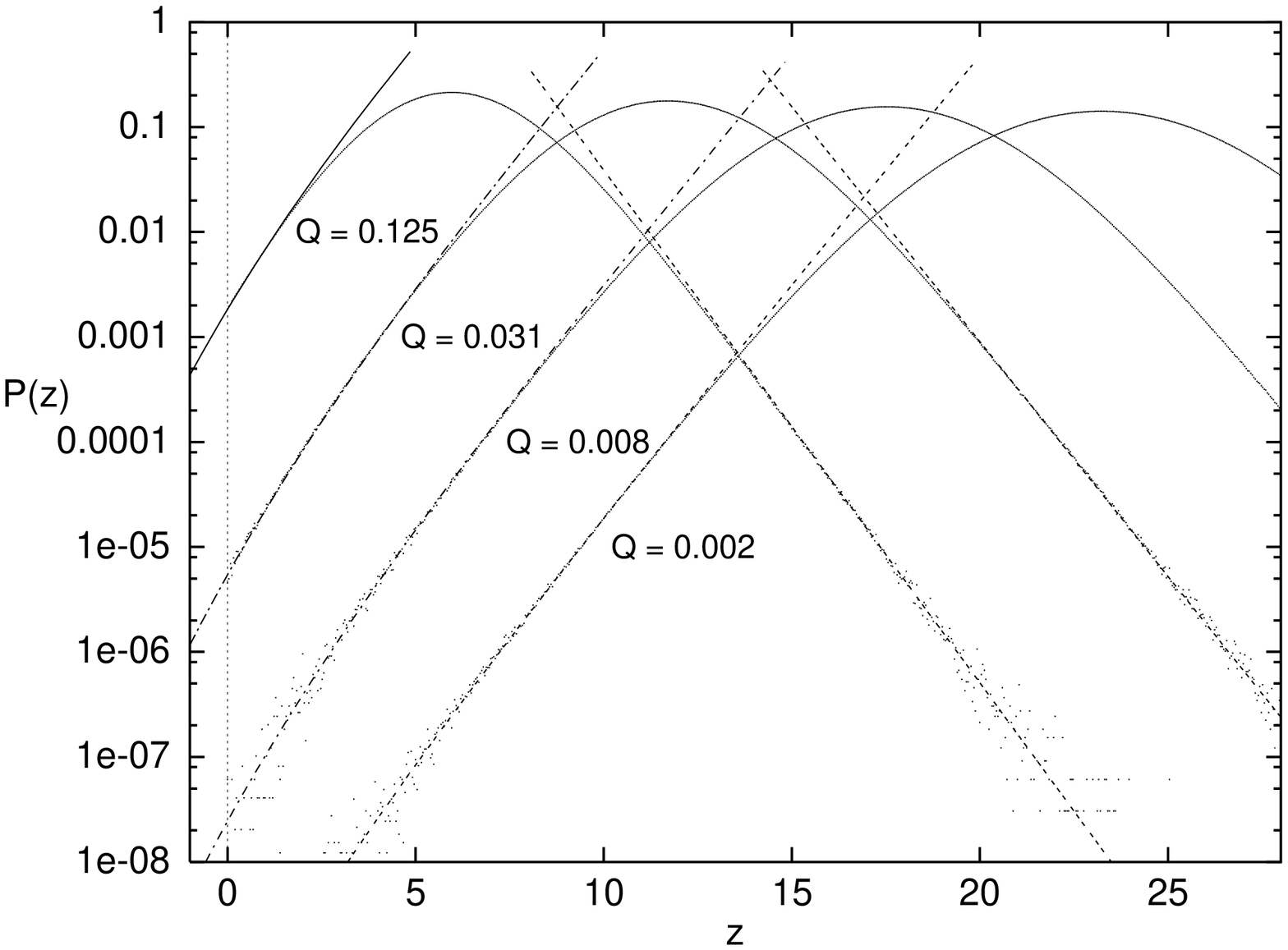}{0.495\textwidth}{ Typical probability distributions
  $\cP(z)$ for the SOS model with one fluctuating surface and a soft
  wall.  The sample size was 64x64 sites, and the hydration
  interaction parameters were set to $U_{hy}=2.0$ and $z_{hy}=4.0$.
  The tails are well represented by a mean field like probability
  distribution $\cP(z)\sim\exp(-|z-\langle z
  \rangle|-Qz-U_{hy}e^{-z/z_{hy}})$. 
  Each tail has been fitted individually.
  In each fit, the only fit parameter was a constant factor. The difference
  to the histograms for the Gau{\ss}ian model, where the exponential
  tails are absent, is obvious.}

Thus, we should distinguish two different types of excitations: (i),
{\em Collective} fluctuation modes that lead to a Gau{\ss}ian shaped
probability distribution and, consequently, to the contact exponent as
given by (\ref{nu2_gaussian_uhy}); and (ii) {\em Isolated} protrusions
of single molecules which are absent in the discrete Gau{\ss}ian model.

Now, let us assume that the tails of the distribution $\cP(l)$ and
hence the contact probability $\cP_{2b}$ is dominated by these rare 
protrusions. The simplest estimate for $\cP_{2b}$ is now given by
\begin{equation}
\label{P2b_single}
  \cP_{2b}^{SOS}  
  \sim \exp[-\ell/l_{sc}],
\end{equation}
which describes the probability for a single protrusion mode to extend over
the distance $\ell$ between the interface and the hard wall. With the
scaling of $\ell$ as given by (\ref{P_lt}) and (\ref{l_t}), we now
find the contact exponent
\begin{equation}
\label{nu2_prime}
  \nu_2 = l_t/l_{sc} = (l_{hy} / l_{sc}) [1 + (l_{pr}/2 l_{hy})^{2}].
\end{equation}
Note that this value depends on the two ratios $l_{hy} /l_{sc}$
and
$l_{hy} / l_{pr}$.

Thus, the Gau{\ss}ian fluctuations and the single protrusion modes lead to a
different asymptotic behavior of the contact probability $\cP_{2b}$. The
true asymptotic behavior will be determined by those fluctuations
which dominate $\cP_{2b}$. This competition leads to another characteristic
length scale $l^{*}_{hy}$ which
separates {\em two  different scaling regimes within the hydration regime}.

Direct comparison of the contact probabilities $\cP_{2b}$ as given by
(\ref{P2b_gaussian}) and (\ref{P2b_single}) leads to the characteristic
length scale 
\begin{equation}
  l^*_{hy} = \frac{l_{pr}}{2}\left( z_{pr} + \sqrt{z_{pr}^2-1} \;\right).
\end{equation}
With $z_{pr}=1.54$, this implies $l^*_{hy}\simeq 2.71\times l_{pr}/2
\simeq 2.09 \;l_{sc}$.

The Gau{\ss}ian fluctuations and the single protrusion modes dominate for
$l_{hy}< l^{*}_{hy}$ and $l_{hy} > l^{*}_{hy}$, respectively. This
leads to the contact exponent
\begin{equation}
\label{nu_final}
\begin{array}{cl}
  \nu_2 =    (l_t/l_{pr})^2 & {\rm for}\quad l_{pr}/2 \le l_{hy} \le
  l^*_{hy} \\
  =  l_t/l_{sc}
  & {\rm for}\quad l_{hy} \ge l^*_{hy}, \rule{0cm}{3ex}
\end{array}
\end{equation}

In Fig.~\ref{zhy1p2b}, MC data for the contact exponent in the SOS model
are shown, together with the theoretical predictions
(\ref{nu_final}). The different values of the contact
exponent $\nu_{2}$ are summarized in Table 1.

\pic{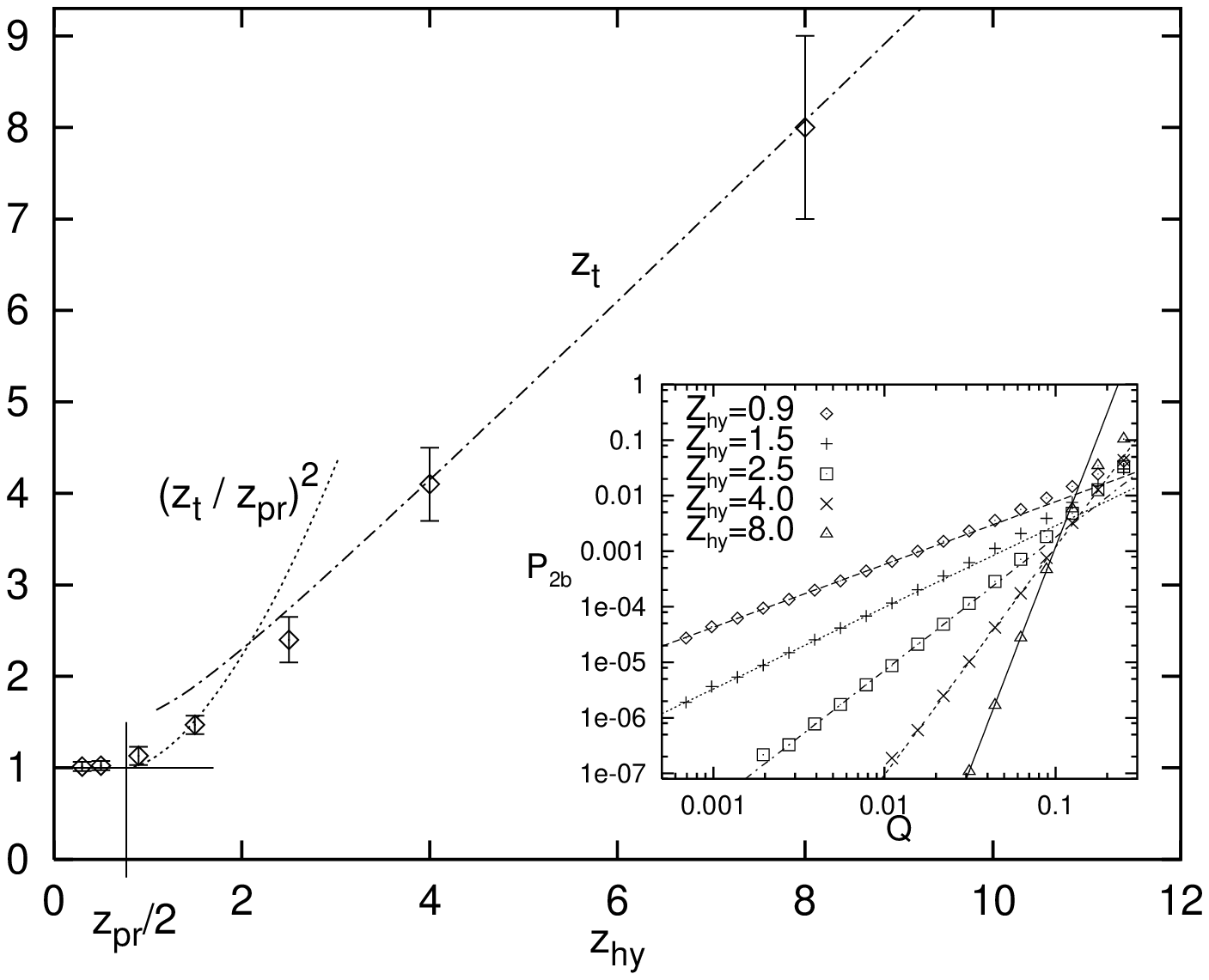}{0.495\textwidth}{
  MC data for a system with one flexible surface and a soft wall. The
  numerical data for the contact exponent $\nu_2$ are
  compared with $\nu_2=1$ in the protrusion regime, and with the 
  functions as given by (\protect\ref{nu_final}) in the hydration
  regime: The   dotted 
  line corresponds to the prediction from Gau{\ss}ian fluctuations,
  $\nu_2=z_t^2/z_{pr}^2$, and the dashed-dotted line to
  $\nu_2=z_t$. Here, we have used $z_{pr}=1.54$. Note that these two
  curves cross at $z_{hy}^*=2.09$. 
{\em Inset:} 
  The MC data for $\cP_{2b}$ are plotted versus the scaled pressure $Q$
  and fitted against $\cP_{2b}\sim Q^{\nu_2}$. Apart from
  $z_{hy}=8.0$, all data cover more than an order of magnitude in
  the pressure range and several orders of magnitude in $\cP_{2b}$.}

\begin{center}
\begin{tabular}{l|c|c|c|}
  Contact & & & \\
  exponent $\nu_2$ & 
  \rb{$l_{hy}<\D\frac{l_{pr}}{2}$} & 
  \rb{$\D\frac{l_{pr}}{2} < l_{hy} < l^{*}_{hy}$} & 
  \rb{$l_{hy}^{*} < l_{hy}$} \\[1.5ex]  \hline
  Discrete Gau-& & \multicolumn{2}{c|}{ }  \rule{0cm}{3.5ex}\\ 
  {\ss}ian Model & \rb{1} & 
  \multicolumn{2}{c|}{ \rb{$(\,l_t\,/\,l_{pr}\,)^2$} } \\[1.5ex] \hline
  Solid--on-- &&&  \rule{0cm}{3.5ex}\\ 
  solid Model   & \rb{1} &  \rb{$(l_t/l_{pr})^2$} & \rb{$l_t/l_{sc}$}
  \\[1.5ex] \hline
\end{tabular}
\end{center}

Table 1. The contact exponent $\nu_{2}$ for the discrete Gau{\ss}ian and the
solid--on--solid  models.


\begin{thebibliography}{10}
\bibitem{footnote1}
 The interface in the 3--dimensional Ising model should undergo a
roughening transition at a characteristic temperature $T = T_r$. Thus, one must
in general distinguish a low--temperature regime with $T < T_r$ from a
high--temperature regime with $T > T_r$. Here, we are only concerned with the
latter regime.
\end{thebibliography}
\end{document}